\documentclass[letterpaper,journal]{IEEEtran}
\usepackage{amsmath,amssymb,amsfonts}
\usepackage{array,booktabs,multirow}
\usepackage{algorithm,algorithmic}
\usepackage{graphicx}
\usepackage{url}
\usepackage{cite}
\usepackage[T1]{fontenc}
\usepackage{newtxmath}
\usepackage{placeins}
\usepackage[most]{tcolorbox}

\newtcolorbox{prompttemplate}[1]{
  breakable,
  sharp corners,
  colback=black!2,
  colframe=black!40,
  colbacktitle=black!10,
  coltitle=black,
  fonttitle=\bfseries\small,
  fontupper=\small,
  title={#1},
  boxrule=0.4pt,
  left=7pt,right=7pt,top=6pt,bottom=6pt,
  before skip=8pt,after skip=8pt
}
\begin{document}
\title{SatMend: LLM Agents for Automated Software Repair in Satellite Constellations with Intermittent Ground Contact}
\author{Pengcheng Luo, Zenghui Zhang, Genke Yang, and Jian Chu%
\thanks{This work was supported by the National Science and Technology Major Project of China under Grant 2025ZD1602400. \emph{(Corresponding authors: Zenghui Zhang and Genke Yang.)}}%
\thanks{Pengcheng Luo, Zenghui Zhang, Genke Yang, and Jian Chu are with Ningbo Artificial Intelligence Institute, Shanghai Jiao Tong University, Ningbo 315000, China, and also with the School of Automation and Intelligent Sensing, Shanghai Jiao Tong University, Shanghai 200240, China, and the Key Laboratory of System Control and Information Processing, Ministry of Education of China, Shanghai 200240, China (e-mail: luopeng69131@sjtu.edu.cn; zenghui.zhang@sjtu.edu.cn; gkyang@sjtu.edu.cn; chujian@niii.com).}%
\thanks{The source code will be made available at \protect\url{https://github.com/luopeng69131/SatMend}.}%
}

\maketitle

\begin{abstract}
Satellite constellations rely on onboard software to sustain services across cooperating spacecraft. Maintaining this software requires handling related component versions and different failure causes while delivering repairs through intermittent ground contact. Automated software repair in satellite constellations remains underexplored, particularly the coordination of repairs across related versions with verified onboard recovery under intermittent ground contact. To address this gap, we propose \emph{SatMend}, a software repair method driven by ground-based large language model (LLM) agents. SatMend organizes related versions into persistent sessions to retain diagnostic context and combines source and runtime evidence to select code repair or state recovery actions. It validates each version separately and submits ready candidates during continued repair, allowing ground preparation to overlap with onboard delivery and verification. Experimental results show that across configurations with 16--64 satellites, SatMend restores 192 of 224 faulty applications within 600~s, compared with 40 for SWE-agent, 77 for Agentless, 124 for RepairAgent, and 115 for SWE-Search. These results demonstrate SatMend's effectiveness in restoring applications and supporting service continuity in satellite constellations with intermittent ground contact.
\end{abstract}

\begin{IEEEkeywords}
Satellite constellations, software repair, large language model agents, intermittent ground contact, service recovery.
\end{IEEEkeywords}

\section{Introduction}
\subsection{Background and Motivation}
\begin{figure}[!t]
  \centering
  \includegraphics[width=\linewidth]{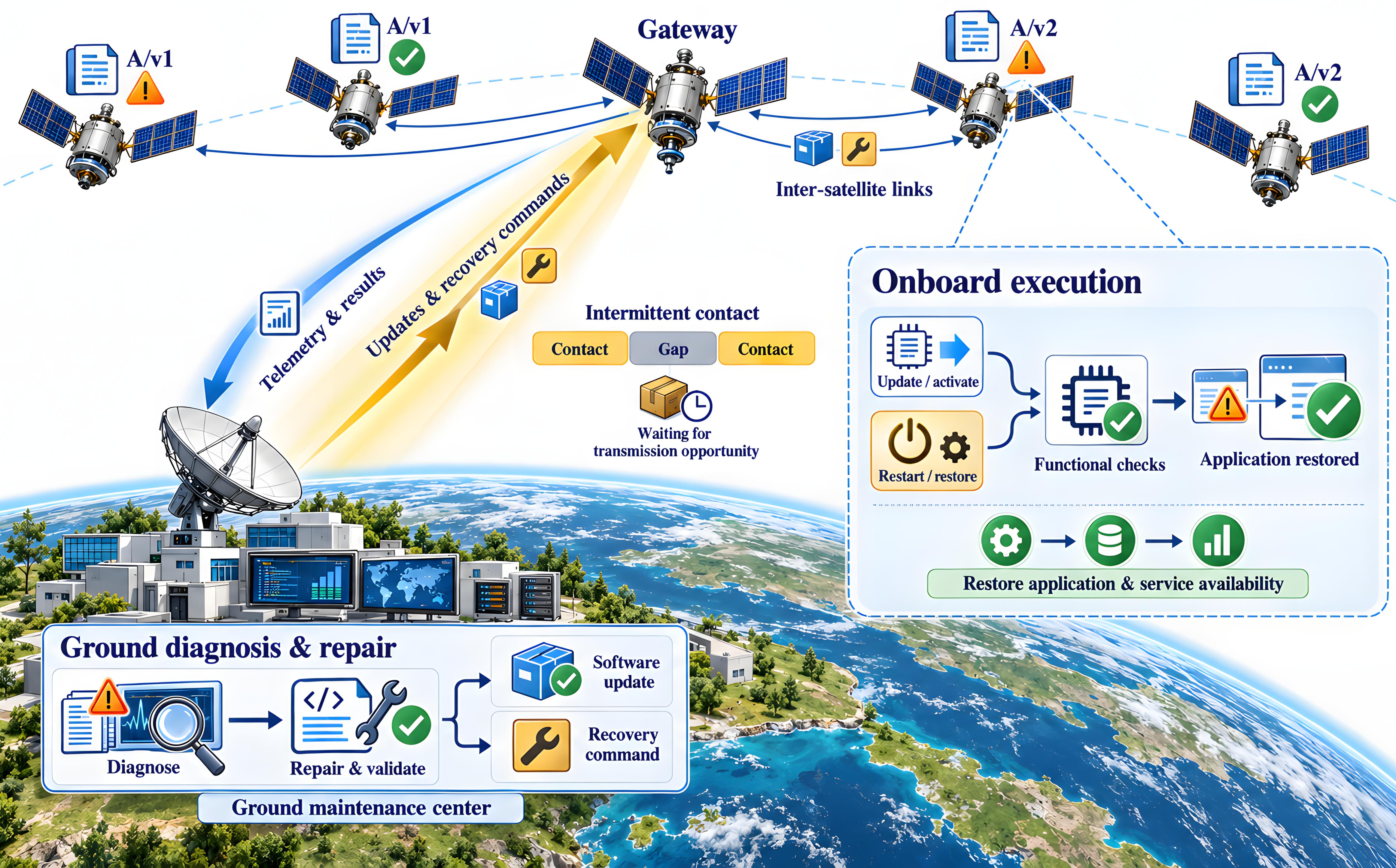}
  \caption{Software maintenance and target recovery in a satellite constellation with intermittent ground contact.}
  \label{fig:intro-background}
\end{figure}

\IEEEPARstart{L}{ow} Earth orbit satellite constellations rely on onboard software for parameter management, file handling, and data forwarding. Software platforms support onboard data processing~\cite{sm_wronkiewicz2026platform}, while crosslinks enable data distribution among satellites~\cite{sm_rodriguezperez2023cache}. Reliable software operation therefore supports both individual spacecraft functions and cooperation across a constellation. A component may be deployed on several satellites, with different versions remaining in operation as updates are introduced. A maintenance task can therefore involve multiple related software instances. Crosslinks also support coordinated maintenance: the National Aeronautics and Space Administration (NASA) has demonstrated file sharing, software updates, and verification through cooperation among satellites during the extended Starling mission~\cite{sm_nasa_starling2025}. In Fig.~\ref{fig:intro-background}, the ground maintenance center sends software updates or commands to restore application state through a gateway satellite. The targets execute these actions and undergo functional checks to confirm application recovery, supporting service continuity across the constellation.

Maintaining these instances requires accounting for both software differences and communication conditions. Versions of the same component share functional context and diagnostic clues, but faults may lie in code, runtime state, or configuration. Telemetry can reveal abnormal behavior~\cite{sm_li2024multichannel,sm_sadr2025uncompromised}, but the appropriate recovery action must still be established for each version. Evidence collection and delivery across targets also share intermittent ground contact, while a failed gateway forwarding application may block downstream maintenance. Analysis and testing based on information already available on the ground can continue during disconnection; new target observations, transfers, and acknowledgments depend on contact opportunities. Even after a candidate repair is ready, it must still be transferred, activated, and checked at each target. Its readiness time affects which delivery opportunities remain available. Recovery by a maintenance deadline thus depends on coordinating repair preparation across related versions with contact availability and target execution.

General-purpose program repair systems use tools and feedback to gather evidence, modify code, and validate candidates~\cite{sm_repairagent2025}, while patch-porting research addresses the applicability of changes across code versions~\cite{sm_llmport2025}. Satellite application updates and file transfer within satellite swarms provide mechanisms for delivering software artifacts~\cite{sm_satelight2025,sm_starling_lft2025}. Automated software repair across satellite constellations, however, remains underexplored. In particular, reusing diagnostic work across related versions must be coordinated with contact-constrained delivery and functional verification at individual targets so that repairs prepared on the ground lead to recovery before the maintenance deadline.

\begin{figure*}[!t]
  \centering
  \includegraphics[width=\linewidth]{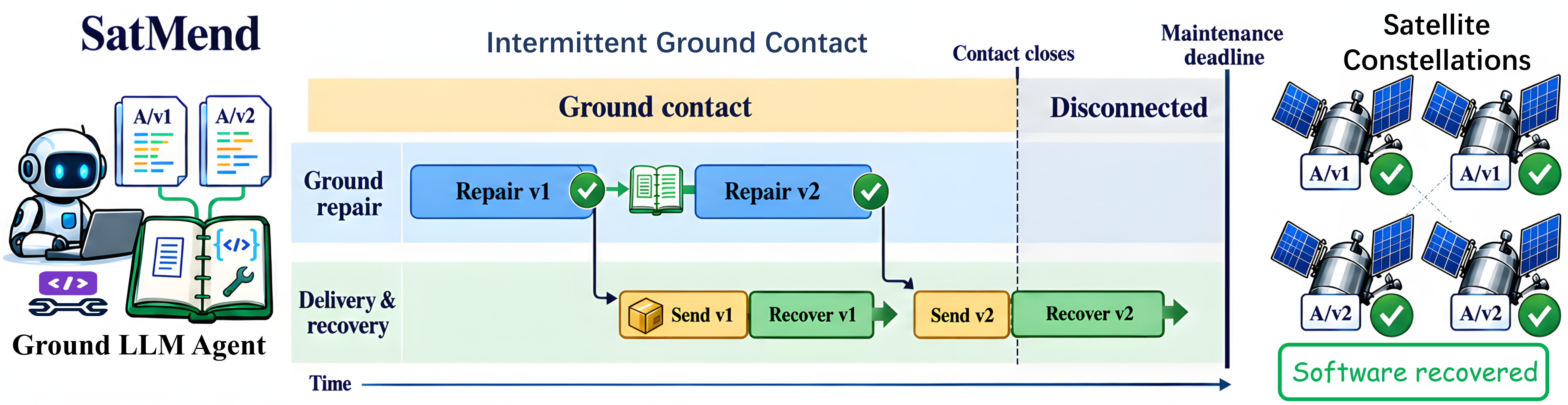}
  \caption{SatMend connects ground repair preparation to onboard recovery through persistent sessions and timely delivery.}
  \label{fig:intro-satmend-concept}
\end{figure*}

A large language model (LLM) agent with tool access can interleave reasoning, actions, and observations, revising its plan as new evidence arrives~\cite{sm_react2023}. In program repair, test feedback and earlier candidates can inform subsequent attempts~\cite{sm_chatrepair2024}; experience retained from feedback can likewise guide later decisions~\cite{sm_reflexion2023}. These capabilities suggest maintaining shared diagnostic context for related versions while validating each against its own evidence. Once a candidate for one version passes validation, it can be submitted while analysis and testing continue for the remaining versions, allowing ground preparation to overlap with onboard execution. Maintaining this context while submitting ready candidates promptly offers a way to turn limited maintenance time into actual application recovery.

To this end, we propose \emph{SatMend}, a software repair method for satellite constellations driven by LLM agents on the ground. SatMend groups versions of shared components into persistent sessions, reuses diagnostic context, and combines source code, available observations, and maintenance dependencies to select actions for code repair or state recovery. It validates candidates separately for each version and submits ready candidates to a shared maintenance executor, which handles delivery, activation, and functional checks for individual targets. Returned execution feedback informs further diagnosis and maintenance. In Fig.~\ref{fig:intro-satmend-concept}, two versions of the same component illustrate this process: repair of a subsequent version overlaps delivery of an earlier candidate, and tasks whose packages and commands have reached the gateway continue onboard after ground contact ends. Persistent sessions and timely delivery thus connect progress across related repairs to target execution, translating completed ground preparation into recovery of applications and their dependent services under intermittent contact.

\subsection{Contributions}
The main contributions of this paper are as follows:
\begin{itemize}
\item We formulate satellite software maintenance under intermittent ground contact, maximizing verified application recovery before a deadline while accounting for version differences, code faults, state anomalies, dependencies, and delivery constraints.

\item We develop SatMend, which reuses diagnostic context across related versions in persistent LLM sessions. It coordinates code repair and state recovery with separate candidate validation and timely delivery, allowing ground preparation to overlap with onboard execution.

\item We evaluate SatMend on a software-in-the-loop testbed built from public F Prime components against six baselines. Across six configurations with 16--64 satellites, it restores 192 of 224 faulty applications within 600~s, compared with the highest baseline total of 124.
\end{itemize}

\subsection{Organization}
Section~II reviews spacecraft software maintenance and LLM-based repair workflows. Section~III presents the system model and maintenance objective. Section~IV describes SatMend's cross-version sessions, evidence-guided repair, and timely delivery. Section~V evaluates recovery performance, contact effects, and method components, followed by a maintenance case study. Section~VI concludes the paper.

\section{Related Work}
Research on spacecraft recovery and on LLM-based software maintenance provides complementary foundations for connecting ground diagnosis, repair preparation, and onboard recovery.

\subsection{Spacecraft Software Recovery and Maintenance}
Spacecraft depend on onboard software to control operations, manage data, and provide mission services. Maintaining these functions involves detecting and diagnosing abnormal behavior, applying recovery actions, and deploying software updates when code changes are needed. Hundman \emph{et al.} use real spacecraft telemetry for anomaly detection with long short-term memory networks and dynamic thresholds~\cite{sm_hundman2018}, while the OPS-SAT benchmark provides labeled telemetry and an evaluation basis for satellite anomaly detection~\cite{sm_opssat2025}. Fault detection, isolation, and recovery (FDIR) links observations to state assessment and corrective actions. Wanninger's knowledge-based FDIR combines events, states, and rule-based reasoning to support recovery in multiple stages~\cite{sm_wanninger2026}. These studies support anomaly identification and recovery execution; selecting an appropriate maintenance action still requires relating current evidence to the software's functional requirements.

When diagnosis must combine telemetry, operational documentation, and procedures, language models offer a way to organize the available information. Campanelli and Hiebl connect document retrieval, telemetry analysis, and command tools to an LLM agent in a simulated satellite operations environment, with operator confirmation before command execution~\cite{sm_campanelli2025}. PHOENIX proposes an architecture combining a small onboard model with ground-based agents and explores the generation of recovery commands within contact windows~\cite{sm_phoenix2026}. LLM-based code repair has also been explored during satellite software development: L{\'o}pez and de Fitero use source code and SonarQube findings to generate corrections~\cite{sm_lopez2024repair}. Applying code repair to deployed applications additionally requires preparing changes for the intended version, delivering them, and verifying their effect at the target.

Software update systems provide part of this execution path. SateLight improves update efficiency and reliability for heterogeneous satellite applications through differential transmission, onboard reconstruction, and layered fault tolerance~\cite{sm_satelight2025}. The Consultative Committee for Space Data Systems (CCSDS) standardizes file delivery and associated file management through the CCSDS File Delivery Protocol (CFDP)~\cite{sm_cfdp2020}. The large file transfer system developed for the Starling swarm uses file chunks and crosslink communication to support transfers among satellites~\cite{sm_starling_lft2025}. These studies provide foundations for spacecraft fault handling and software delivery. Automated repair of deployed satellite applications remains underexplored, particularly across related software versions whose recovery must be achieved under intermittent ground contact.

\subsection{LLM-Based Software Repair and Maintenance Workflows}
Automated program repair aims to restore intended software behavior by identifying and correcting defects in code. LLM-based approaches use code context to propose candidate fixes and execution feedback to guide validation and further refinement. Xia \emph{et al.} compare pretrained language models and strategies for code generation in program repair, examining how context and candidate generation affect repair outcomes~\cite{sm_xia2023repair}. ChatRepair incorporates candidate patches and test feedback into a continuing conversation, allowing later attempts to use information from earlier successes and failures~\cite{sm_chatrepair2024}. ThinkRepair further combines repair reasoning examples with test feedback to guide candidate construction~\cite{sm_thinkrepair2024}. Using these inputs throughout repair requires a workflow for obtaining relevant evidence and deciding which actions to take next.

At repository scale, this workflow must locate relevant implementations and coordinate inspection, editing, and testing. SWE-bench evaluates modifications to real repositories in their execution environments, connecting issue descriptions, repository context, and tests~\cite{sm_swebench2024}. SWE-agent supports interactive repair through browsing, editing, and testing interfaces designed for agents~\cite{sm_sweagent2024}, while AutoCodeRover uses program structure and code search to obtain relevant context~\cite{sm_autocoderover2024}. Control strategies also differ: Agentless separates localization, repair, and validation into stages~\cite{sm_agentless2025}; RepairAgent uses a finite-state machine to guide evidence collection and candidate validation~\cite{sm_repairagent2025}; and SWE-Search combines tree search with feedback to evaluate candidate trajectories~\cite{sm_swesearch2025}. These methods organize evidence and actions within a repair task; components maintained in several versions also offer opportunities to reuse the resulting diagnosis and changes across tasks.

Patch-porting research addresses one form of this reuse by adapting a fix from one code context to another. PPatHF uses the code before and after a reference patch, together with the corresponding target function, to adapt fixes to a forked project~\cite{sm_ppathf2024}. LLMPort supports patch porting across files through task decomposition and self-correction~\cite{sm_llmport2025}. These methods exploit shared code while adapting changes to each target version. When several versions raise alarms together, shared context and evidence for individual versions must also be organized as diagnosis proceeds. Once a candidate is validated, its contribution to recovery further depends on execution at the deployed targets.

Evaluating this final step requires observing the running system after maintenance actions. AIOpsLab supports evaluation of agent diagnosis, operations, and recovery through interactive cloud environments and fault lifecycles~\cite{sm_aiopslab2025}. ITBench likewise uses executable operational tasks to assess whether agents achieve system maintenance goals~\cite{sm_itbench2025}. These advances provide tools for constructing, adapting, and evaluating repairs. In satellite constellations, their contribution to recovery additionally depends on how diagnostic work carries across software versions and whether delivery and target verification can complete within the maintenance deadline despite intermittent ground contact.

\section{System Model and Problem Formulation}
\label{sec:system-model}
Software maintenance in a satellite constellation connects diagnosis and candidate preparation on the ground to execution at onboard targets. Shared components allow related applications to reuse part of the repair work, while differences in version, runtime state, and ground contact affect whether a repair takes effect. This section first defines software deployments, available information, and maintenance actions, then models execution constraints and the objective of recovery before a deadline.

\subsection{Satellite Software Maintenance Architecture}
\label{sec:maintenance-architecture}
In Fig.~\ref{fig:intro-background}, a software component can be deployed on multiple satellites and exist in different source versions. Relating components, versions, and deployed replicas identifies which repair tasks share a common context and which targets must receive each candidate.

Let $\mathcal V$ be the set of satellites involved in a maintenance operation and $\mathcal A$ the set of deployed applications. Application $a\in\mathcal A$ runs on satellite $p(a)\in\mathcal V$ and has component type $\kappa(a)$ and source version $\nu(a)$. Multiple applications may share both a component type and a source version, or implement the same component using different versions. Source-version identifiers associate production code with deployed targets; the image actually running at a target is checked before delivery.

The ground maintenance platform holds production sources for each version, public functional contracts, and the deployment inventory. It runs LLM agents, build tools, and candidate tests. Each satellite hosts target applications and an independent update executor that receives packages, activates target programs, and performs functional checks. A candidate that passes ground validation can serve multiple replicas of its applicable version. Each replica still requires individual delivery and verification.

Maintenance traffic enters the interconnected satellites through gateway $g\in\mathcal V$. Its forwarding application $\rho\in\mathcal A$ provides the forwarding function needed to maintain downstream targets, so its state affects their accessibility. The independent update executor is deployed separately from the applications being repaired. The ground can therefore maintain the gateway itself during contact even when its forwarding application fails. This architecture connects shared repair preparation to execution at individual targets and makes application state and service dependencies relevant to subsequent maintenance decisions.

\subsection{Fault States, Observations, and Maintenance Actions}
\label{sec:maintenance-actions}
A service alarm may originate from an application's own code or runtime state, or from a failed upstream dependency. Recovery therefore requires relating source code, functional contracts, and observations to decisions about code changes, state recovery, or restoration of a prerequisite application. The model distinguishes target functionality from the information already available on the ground.

Let $h_a(t)\in\{0,1\}$ denote the health of application $a$ at time $t$. It equals one when the application meets its specified functional requirements and passes the checks of normal operation, and zero otherwise. Code faults require a candidate built for the corresponding source version; anomalies in runtime state or persistent configuration may be resolved through supported operations for restart or configuration restoration. Application state also affects other targets through service dependencies. Each application is associated with a service, and a public directed acyclic graph $\mathcal D$ describes the applications on which that service depends. A healthy application can therefore raise a service alarm when an upstream application is unavailable.

At the start of maintenance, the ground receives deployment and version inventories, cached alerts, production sources, functional contracts, dependencies, and a contact calendar. Let $\mathcal B_0$ be the set of targets appearing in the cached alerts. It can include both faulty applications and applications affected by dependencies; their maintenance needs must be determined from subsequent evidence. New runtime observations must be obtained through the maintenance links. During disconnection or while a request remains incomplete, cached information retains its original observation time.

Let $\mathcal H_t$ denote the maintenance information obtained by time $t$, including initial inputs, completed ground operations, submitted target requests, and feedback actually received. At decision time $t_k$, maintenance policy $\pi$ selects an action from this history:
\begin{equation}
u_k=\pi(\mathcal H_{t_k}).
\label{eq:maintenance-policy}
\end{equation}

Actions include source inspection and editing on the ground, candidate building and validation, target inspection, state recovery, software delivery, and execution-status queries. Ground analysis and building can continue with existing information, while obtaining new observations or acting on a target requires suitable communication conditions. Build diagnostics, functional results, and target receipts become available to subsequent decisions, allowing maintenance to adapt to actual outcomes.

The policy thus updates its assessment of alarm causes and execution progress using only information already obtained. The initial set of directly faulty applications and the complete target state are used to evaluate recovery outcomes. Individual actions must also satisfy the contact, path, and resource conditions below.

\subsection{Maintenance Execution under Contact Constraints}
\label{sec:maintenance-execution}
Communication requirements vary across maintenance stages: source analysis and candidate building use information held on the ground, whereas target inspection and uplink delivery require contact. After a request reaches the satellites, gateway forwarding state and shared crosslink resources affect further execution. The model therefore connects ground contact, geometric reachability, resource capacity, and version applicability to the conditions under which a repair takes effect.

Let $T$ be the maintenance deadline and $\mathcal W_G\subseteq[0,T]$ the known union of ground contact windows. Contact availability between the ground and gateway is
\begin{equation}
c_G(t)=\mathbf 1\{t\in\mathcal W_G\}.
\label{eq:contact-availability}
\end{equation}
New inspection requests, uplink commands, and downlink feedback must complete their respective byte transfers during contact. A request reaching its destination does not imply that the ground has received its result; the return transfer is still subject to contact availability and capacity. Local source inspection, editing, and build tests do not require $c_G(t)=1$.

For application $a$ on satellite $p(a)$, let $\ell_{g,p(a)}\in\{0,1\}$ indicate whether the gateway can reach the host satellite within the allowed crosslink range without obstruction by the Earth. The current maintenance architecture uses fixed geometry among satellites sharing an orbit, while path availability varies with the gateway forwarding application:
\begin{equation}
\chi_a(t)=
\begin{cases}
1, & p(a)=g,\\
\ell_{g,p(a)}h_\rho(t), & p(a)\ne g.
\end{cases}
\label{eq:maintenance-path}
\end{equation}
The independent executor handles access to the gateway itself. Access to other satellites requires both geometric reachability and a functioning gateway forwarding application. Restoring this application can therefore change which subsequent maintenance actions are feasible. The path state $\chi_a(t)$ belongs to the execution environment; the ground learns about it through actual access attempts and feedback.

Let $B_G^\uparrow$, $B_G^\downarrow$, and $B_I$ be the uplink, downlink, and shared crosslink capacities, respectively, in bit/s. Write $x_G^d(t)$ for the executor's actual ground-link transfer rate in direction $d\in\{\uparrow,\downarrow\}$. Let $\mathcal J_I(t)$ be the active crosslink transfers, with actual rate $x_j(t)$ for transfer $j$. Capacity constraints are
\begin{equation}
0\le x_G^d(t)\le B_G^d c_G(t),\qquad
\sum_{j\in\mathcal J_I(t)}x_j(t)\le B_I.
\label{eq:maintenance-capacity}
\end{equation}

The executor advances transfers according to their byte requirements, resource queues, and contact interruptions. The ground also has $K$ repair slots. Model interactions, builds and tests, failed attempts, transfers, and target operations all follow a common maintenance timeline. Concurrent or overlapping work is accounted for using its actual start and end times, and sessions share the executor and communication resources.

Delivery additionally requires candidate validation and version applicability. For candidate $r$, let $v(r)$ be its source version and $\gamma(r)\in\{0,1\}$ its result of ground validation. Delivery to application $a$ requires $\gamma(r)=1$ and $\nu(a)=v(r)$, and binds the package to the base image actually running at that target. Validation applies to specific candidate contents; further source changes require validation of the new candidate.

An eligible candidate must complete uplink transfer to the gateway, forwarding to each target, activation, and functional verification. The gateway can cache a complete package for reuse by compatible targets, but execution and verification are recorded separately for each target. Once a package and its target commands have reached the gateway in full, crosslink delivery and activation can continue while the ground is disconnected. Partial uploads require the remaining uplink transfer, and confirmation on the ground requires the corresponding feedback. Candidate readiness, target recovery, and ground confirmation therefore have distinct completion times.

\subsection{Recovery Objective}
\label{sec:recovery-objective}
A single validated candidate may serve several targets, but recovery before the deadline also depends on delivery and target execution. The primary objective therefore counts initially faulty applications that are actually restored. Service availability captures how their recovery affects dependent functions.

For maintenance configuration $c$, define the initial set of directly faulty applications as $\mathcal F_c=\{a:h_{c,a}(0)=0\}$. Let $h_{c,a}^\pi(t)$ denote application health under policy $\pi$. An initially faulty application counts as recovered only after it passes the required functional checks and checks of normal operation at its actual target. The recovered application count at the deadline is
\begin{equation}
N_c^\pi(T)=\sum_{a\in\mathcal F_c}h_{c,a}^\pi(T).
\label{eq:recovered-applications}
\end{equation}
Each initially faulty application is a separate recovery unit. Replicas of the same version are counted individually when restored. Ground validation, package upload, and termination of an agent session are intermediate milestones in the maintenance process.

Let $\Pi_{\mathrm{feas}}$ contain policies that act on available information and respect contact, resource, version compatibility, and verification constraints. The maintenance objective is
\begin{equation}
\max_{\pi\in\Pi_{\mathrm{feas}}}N_c^\pi(T).
\label{eq:maintenance-objective}
\end{equation}
The benefit of a candidate depends on how many compatible targets complete recovery before the deadline. Multiple versions occupy limited repair slots, and multiple targets share delivery resources. Repair preparation and target execution therefore both consume the maintenance budget; failed builds and incomplete attempts also reduce the time available for subsequent recovery.

Application recovery can also restore dependent services. For a fixed policy, use each application identifier to denote its associated service, and let $\operatorname{Anc}_{\mathcal D_c}(a)$ be the set of all upstream application dependencies of $a$ in configuration $c$. Service availability is
\begin{equation}
A_{c,a}(t)=\prod_{b\in\{a\}\cup\operatorname{Anc}_{\mathcal D_c}(a)}h_{c,b}(t).
\label{eq:service-availability}
\end{equation}
This value is one when the application and all its upstream dependencies are healthy, and zero otherwise. Restoring a prerequisite application can therefore unblock multiple dependent services. Application counts measure completed maintenance work, while service recovery and interruption time capture its operational effects. Improving these outcomes requires coordinating repair preparation across versions with target execution under contact constraints so that completed repairs take effect before the deadline.

\begin{figure*}[!t]
  \centering
  \includegraphics[width=\linewidth]{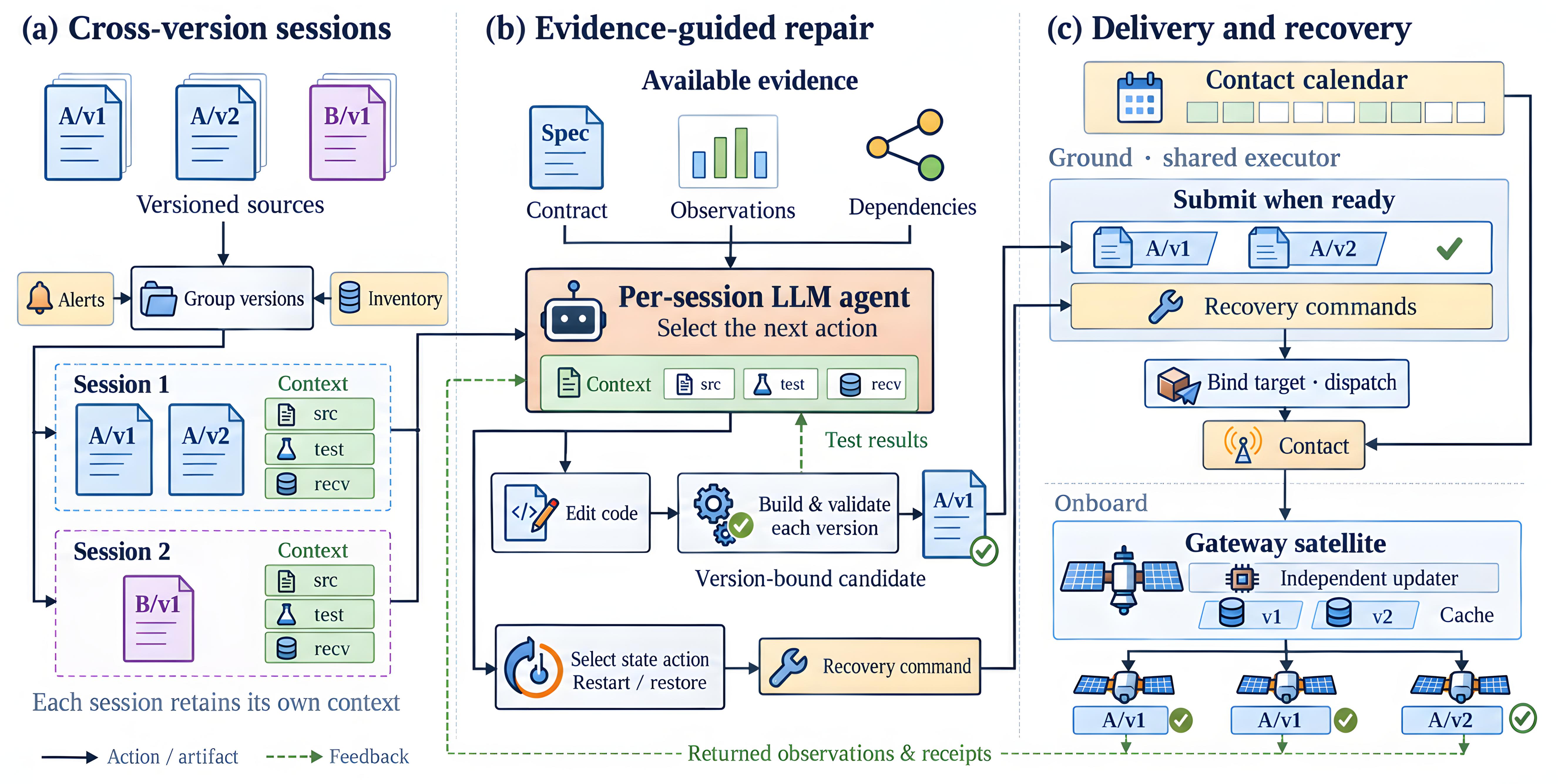}
  \caption{SatMend framework: (a) cross-version sessions; (b) evidence-guided repair and validation for each version; (c) timely delivery and target recovery.}
  \label{fig:satmend-framework}
\end{figure*}

\section{SatMend Design}
\label{sec:satmend}
SatMend connects repair preparation across source versions to target execution under contact constraints, aiming to restore more applications before the maintenance deadline. Persistent sessions retain the analysis of related versions, source code and execution evidence guide maintenance actions, and delivery requests follow repair progress. These mechanisms allow ground repair preparation and onboard recovery to proceed together.

\subsection{Framework Overview}
\label{sec:satmend-overview}
Versions of a shared component provide common context for diagnosis, but each version still requires its own candidate validation. Once a candidate is ready, the remaining contact time and maintenance path determine whether it can restore its targets. SatMend connects these stages through the three modules in Fig.~\ref{fig:satmend-framework}: cross-version sessions in (a), evidence-guided repair and validation in (b), and timely delivery with recovery feedback in (c).

In Fig.~\ref{fig:satmend-framework}(a), the ground identifies versions to inspect from cached alerts and the deployment inventory, then organizes persistent sessions around shared components. Each session retains its source analysis, test results, and execution receipts so that subsequent work on related versions can use the established context. Together with public functional contracts, available observations, and service dependencies, this context supports the agent's action selection in Fig.~\ref{fig:satmend-framework}(b). Code faults lead to editing, building, and validation for each version, while anomalies in state or configuration are addressed through the corresponding restart or restoration operations.

In Fig.~\ref{fig:satmend-framework}(c), the shared maintenance executor handles the candidates and recovery operations prepared by these sessions. Validated candidates can be submitted as repair progresses. After target-version checks, packages reach the gateway through available contact opportunities, and onboard executors handle forwarding, activation, and functional verification for individual targets. The gateway cache allows compatible targets to reuse received packages. Available observations and execution receipts return to the originating session, supporting further diagnosis, restoration of maintenance prerequisites, or revised delivery requests. Ground preparation of subsequent versions can thus overlap onboard execution of tasks already received in full, turning progress within sessions into recovery at deployed targets.

\subsection{Cross-Version Repair Sessions}
\label{sec:cross-version-sessions}
Versions of a component often share functional contracts, interfaces, and parts of their implementation. Starting a new session for each version would repeat the work of reading and interpreting this material. SatMend groups repair work using public component relationships so that diagnostic findings and validation feedback remain available as a session proceeds across related versions. The assignment also defines which sources each session may modify, allowing repair slots to work on different version groups in parallel.

The cached alert set $\mathcal B_0$ identifies targets to inspect. The component and version mappings in Section~\ref{sec:maintenance-architecture} then define, for component type $j$,
\begin{equation}
\begin{aligned}
\mathcal G_j&=\{\nu(a):a\in\mathcal B_0,\ \kappa(a)=j\},\\
\mathcal A_v&=\{a\in\mathcal B_0:\nu(a)=v\}.
\end{aligned}
\label{eq:component-version-groups}
\end{equation}
Here, $\mathcal G_j$ contains the source versions of component $j$ implicated in the alerts, and $\mathcal A_v$ contains the alerted targets associated with version $v$. This organization preserves the relation between components, versions, and targets: components connect diagnostic context, versions identify candidate repairs, and targets identify where recovery must occur. Subsequent diagnosis determines whether an alert originates in the application itself.

Let $\mathcal S_k$ be the set of versions assigned to repair slot $k$, initially empty. SatMend processes component groups in decreasing order of $|\mathcal G_j|$ and assigns each group in full to a slot with the fewest assigned versions:
\begin{equation}
\begin{aligned}
k^*&\in\arg\min_{1\le k\le K}|\mathcal S_k|,\\
\mathcal S_{k^*}&\leftarrow\mathcal S_{k^*}\cup\mathcal G_j.
\end{aligned}
\label{eq:session-assignment}
\end{equation}
Related versions of a component therefore remain in one slot, while a slot may handle several component groups. Version counts distribute work across slots; the session determines its repair order using dependencies, observations, and execution progress.

Each nonempty slot starts one persistent LLM session. As it works through the assigned versions, the session retains the source code it has read, functional analysis, runtime observations, validation results, and delivery feedback. Later versions can then use the component context established earlier. Source modifications are restricted to versions in $\mathcal S_k$; public information about other versions and accessible targets remains available for inspection. Sessions coordinate through the shared target status and execution records available to the ground, while retaining their own interaction contexts. This organization gives each session a defined version scope and target list. Continued interaction then combines shared component context with evidence for individual versions to determine subsequent maintenance actions.

\subsection{Evidence-Guided Repair and Validation}
\label{sec:repair-validation}
Persistent sessions support analysis across related versions, but similar alerts can require different maintenance actions. With service dependencies, a downstream alarm may result from a failed prerequisite, so editing downstream code may leave the cause unresolved. SatMend combines public functional contracts, source code, and available runtime evidence to identify what to inspect and how to restore functionality, then revises that assessment using execution results.

A session first uses functional contracts and interfaces to establish the operations that a component should provide. It checks the prerequisite applications implicated by the alerts through the service dependencies, then compares observed behavior with expected functionality using the relevant sources, available runtime observations, and accessible healthy instances. Access results, functional checks, and telemetry remain in the session context as evidence for deciding whether to inspect further or attempt recovery.

When the evidence points to production logic, the agent modifies its assigned source version and uses the diagnostic findings to inspect corresponding implementations in other versions of the component. When the evidence indicates an anomaly in runtime state or configuration, it selects a supported restart or configuration restoration operation. If a prerequisite application blocks maintenance access or service functionality, the session can restore that prerequisite before continuing with downstream targets. Related versions thus share diagnostic context, while each action remains grounded in the relevant code and runtime evidence.

A code change produces a candidate $r$ associated with a specific source version. The shared maintenance executor performs a native build, functional checks, and checks of normal operation; $\gamma(r)=1$ only when all checks pass. The validation record identifies both $v(r)$ and the candidate contents. An unchanged source snapshot can reuse its validation result, whereas further edits require validation of the new candidate. Analysis can therefore carry across versions while delivery eligibility remains tied to the contents actually checked for each version.

Build diagnostics and overall functional-check results return to the originating session to assess the repair hypothesis. If validation fails, the agent revises the candidate and checks it again, with these attempts consuming the same maintenance budget. State recovery is confirmed through operations and functional checks at the actual target. Validated candidates and their applicable targets proceed to delivery, while unresolved work continues in the same session using the new feedback.

\subsection{Timely Delivery and Recovery Feedback}
\label{sec:delivery-feedback}
After one candidate passes validation, preparing later versions may consume the remaining contact opportunity. SatMend interleaves delivery requests with continued repair so that ready versions can enter delivery while others remain under analysis or validation. The agent chooses when to submit using the results obtained, target scope, and maintenance conditions; the shared maintenance executor then carries out the request.

A delivery request identifies the candidate version and target applications. The executor checks the validation record against the current source contents, rebuilds and tests when necessary, and obtains the image version actually running at each target. Once source-version applicability is established, targets with the same base image can share a package bound to that image, with target commands specifying its recipients. A complete package can be reused from the gateway cache, while execution tasks and verification records remain separate for each target.

The executor tracks gateway upload progress separately from execution at individual targets. If an uplink transfer is interrupted, the gateway retains the received portion so that a subsequent delivery request can resume from the recorded progress. Once the complete package and target commands reach the gateway, forwarding, activation, and verification proceed under the path and resource constraints in Section~\ref{sec:maintenance-execution}. These tasks can continue during ground disconnection. When downstream maintenance is waiting for gateway forwarding, restoring the gateway application allows existing pending tasks to proceed.

Delivery feedback returns to the originating session. A validation failure supports further candidate revision; an unreached request, partial upload, or unavailable path helps the agent decide whether to resubmit or restore a maintenance prerequisite. Completed target checks update the remaining work. The executor reuses an existing task for the same target and package. The shared workflow also retains validated, already requested deliveries whose targets have not entered the onboard execution queue. After all sessions end, it retries these pending deliveries once within the remaining budget.

Subsequent decisions use feedback already returned, and the ground updates its known execution state when the corresponding receipts arrive. Meanwhile, tasks fully received onboard continue under the execution conditions. Persistent sessions preserve repair context across related versions, and timely submission lets completed preparation use available maintenance paths. Together, they support target recovery before the deadline.

Algorithm~\ref{alg:satmend} summarizes this workflow. LLM tool interactions within each persistent session select maintenance actions, while the shared maintenance executor handles candidate validation and execution at deployed targets.

\begin{algorithm}[!t]
\caption{SatMend Software Repair}
\label{alg:satmend}
\begingroup
\renewcommand{\algorithmicrequire}{\textbf{Input:}}
\renewcommand{\algorithmicensure}{\textbf{Output:}}
\begin{algorithmic}[1]
\REQUIRE Public inventory, cached alerts $\mathcal B_0$, versioned sources and functional contracts, service dependencies, contact calendar, $K$ repair slots, deadline $T$, LLM agent, and shared maintenance executor.
\STATE Form component groups $\mathcal G_j$ and target sets $\mathcal A_v$.
\STATE Initialize $\mathcal S_1,\ldots,\mathcal S_K\leftarrow\emptyset$.
\STATE Assign each $\mathcal G_j$, in decreasing size, to a slot with the fewest versions.
\FOR{each nonempty $\mathcal S_k$ \textbf{in parallel}}
  \STATE Start a persistent session with public inputs and assigned targets.
  \WHILE{the session is active and time remains before $T$}
    \STATE Select the next permitted action using the session context.
    \IF{the action requests candidate delivery}
      \STATE Validate the current candidate and bind compatible targets.
      \IF{validation and target binding succeed}
        \STATE Submit package transfer and per-target execution tasks.
      \ENDIF
    \ELSE
      \STATE Execute the selected inspection, edit, test, or recovery action.
    \ENDIF
    \STATE Append returned feedback to the session context.
  \ENDWHILE
\ENDFOR
\STATE Retry already requested, unqueued deliveries once within the remaining budget.
\STATE Continue advancing queued onboard tasks until $T$.
\ENSURE Candidates bound to source versions, maintenance execution records, and available target-verification and delivery feedback.
\end{algorithmic}
\endgroup
\end{algorithm}

\section{Performance Evaluation}
\label{sec:evaluation}

\subsection{Experimental Setup}
\label{sec:experimental-setup}
\subsubsection{Experimental Environment and Maintenance Scenarios}
The experiments use a software-in-the-loop testbed that combines native flight-software components with contact-constrained maintenance execution. Ground LLM agents inspect and modify sources, build candidates, and issue maintenance requests. Isolated native C++ processes represent onboard applications, while a shared maintenance executor advances communication and target operations under the contact and resource constraints.

The software is derived from public NASA F Prime component sources~\cite{sm_fprime2018}. Table~\ref{tab:software-components} characterizes the eight variants by component, maintenance condition, and recovery operation. Four code variants cover three component families, with FileManager represented by two implementations with different code contexts. Two further PrmDb variants exercise transient state and persistent configuration anomalies while retaining correct source code, and two healthy components provide controls for normal operation. The three code-fault families and two state-recovery cases form five repair units; the two FileManager implementations share one repair unit. Deployed replicas count separately toward application recovery, while the fault inventory remains these five repair units.

\begin{table*}[!t]
\centering
\caption{Software variants, maintenance conditions, and recovery operations.}
\label{tab:software-components}
\begingroup
\small
\renewcommand{\arraystretch}{1.18}
\setlength{\tabcolsep}{5pt}
\begin{tabular}{@{}
  >{\raggedright\arraybackslash}m{\dimexpr0.17\textwidth-1.5\tabcolsep\relax}
  >{\centering\arraybackslash}m{\dimexpr0.09\textwidth-1.5\tabcolsep\relax}
  >{\raggedright\arraybackslash}m{\dimexpr0.48\textwidth-1.5\tabcolsep\relax}
  >{\raggedright\arraybackslash}m{\dimexpr0.26\textwidth-1.5\tabcolsep\relax}@{}}
\toprule
\textbf{Component} & \textbf{Variants} & \textbf{Maintenance condition} & \textbf{Recovery operation} \\
\midrule
PrmDb & 1 & Code faults in parameter staging, commit publication, and persistence & Repair source, build, and deploy \\
FileManager & 2 & Code faults in file move/append direction and directory-list totals & Repair source, build, and deploy \\
FileWorker & 1 & Code faults in resumed-transfer byte counts and completion reporting & Repair source, build, and deploy \\
\addlinespace[3pt]
PrmDb & 1 & Transient runtime-state anomaly with correct source code & Restart the same image \\
PrmDb & 1 & Persistent configuration mismatch with correct source code & Restore configuration \\
\addlinespace[3pt]
DpCatalog, Health & 2 & Healthy controls for normal functionality & Preserve normal operation \\
\bottomrule
\end{tabular}
\endgroup
\end{table*}

Table~\ref{tab:satellite-environment} lists the satellite and communication settings. The maintained satellites form a compact group in one orbital plane and receive ground maintenance through a shared gateway. The gateway forwarding application's actual functional state controls downstream access, while an independent update executor allows the gateway itself to be maintained. A public service dependency graph determines service availability from component health. Ground contact is generated from orbital geometry and station visibility; transfer progress additionally depends on shared link capacities and crosslink reachability. Complete packages and target commands received by the gateway can continue through onboard delivery, activation, and verification during ground disconnection.

\begin{table}[!t]
\centering
\caption{Satellite maintenance environment.}
\label{tab:satellite-environment}
\begingroup
\small
\renewcommand{\arraystretch}{1.18}
\setlength{\tabcolsep}{3pt}
\begin{tabular}{@{}
  >{\raggedright\arraybackslash}m{\dimexpr0.39\columnwidth-\tabcolsep\relax}
  >{\raggedright\arraybackslash}m{\dimexpr0.61\columnwidth-\tabcolsep\relax}@{}}
\toprule
\textbf{Parameter} & \textbf{Setting} \\
\midrule
\multicolumn{2}{@{}l}{\textit{Orbit and visibility}} \\
Orbit & Circular; 650 km altitude; $97.98^\circ$ inclination \\
Ground stations & Kiruna, Redu, and Kourou \\
Contact generation & Geometric visibility; $20^\circ$ minimum elevation; 1 s sampling \\
Satellite deployment & One orbital plane; 2 km between adjacent satellites \\
Crosslink reachability & Maximum distance 300 km; no obstruction by the Earth \\
\addlinespace[3pt]
\multicolumn{2}{@{}l}{\textit{Communication and maintenance}} \\
Ground capacities & 256 kbit/s uplink; 1 Mbit/s downlink \\
Crosslink capacity & 1 Mbit/s shared across transfers \\
Software deployment & Two applications per satellite; one directly faulty \\
Maintenance ingress & One gateway with an independent update executor \\
Maintenance deadline & 600 s \\
\bottomrule
\end{tabular}
\endgroup
\end{table}

The scenario library combines three satellite counts, two initial gateway states, two state anomaly fractions, and four deployment layouts, yielding 48 primary configurations, each with a paired contact calendar. The primary and paired calendars have 90 s and 270 s of initial remaining contact, respectively, with subsequent windows retained. The model-based comparison uses the six configurations listed in Table~\ref{tab:maintenance-configurations}. Each satellite hosts two applications, one with a direct anomaly, giving 448 applications and 224 directly faulty applications across these configurations: 140 with code faults and 84 with state or configuration anomalies. The state anomaly fraction is the latter category's share of initially faulty applications. These runs evaluate known faults under different deployment and contact combinations.

Configuration identifiers specify satellite count and initial gateway state: the number after S is the satellite count, and N (normal) or F (faulty) denotes the initial state of the gateway forwarding application. Thus, S16-N has 16 satellites and a normal forwarding application. The contact windows in Table~\ref{tab:maintenance-configurations} cover the full maintenance interval; the initial 90 s is the time remaining in the current pass when maintenance begins.

\begin{table*}[!t]
\centering
\caption{Evaluated maintenance configurations.}
\label{tab:maintenance-configurations}
\begingroup
\small
\renewcommand{\arraystretch}{1.16}
\setlength{\tabcolsep}{4pt}
\begin{tabular}{@{}
  >{\raggedright\arraybackslash}m{\dimexpr0.13\textwidth-1.714286\tabcolsep\relax}
  >{\centering\arraybackslash}m{\dimexpr0.09\textwidth-1.714286\tabcolsep\relax}
  >{\centering\arraybackslash}m{\dimexpr0.12\textwidth-1.714286\tabcolsep\relax}
  >{\centering\arraybackslash}m{\dimexpr0.15\textwidth-1.714286\tabcolsep\relax}
  >{\centering\arraybackslash}m{\dimexpr0.17\textwidth-1.714286\tabcolsep\relax}
  >{\centering\arraybackslash}m{\dimexpr0.125\textwidth-1.714286\tabcolsep\relax}
  >{\centering\arraybackslash}m{\dimexpr0.215\textwidth-1.714286\tabcolsep\relax}@{}}
\toprule
\textbf{Configuration} & \textbf{Satellites} & \shortstack{\textbf{Initial}\\\textbf{gateway state}}
& \shortstack{\textbf{Applications}\\\textbf{with code faults}}
& \shortstack{\textbf{Applications}\\\textbf{with state anomalies}}
& \shortstack{\textbf{State anomaly}\\\textbf{fraction}}
& \shortstack{\textbf{Contact windows}\\\textbf{within 600 s (s)}} \\
\midrule
S16-N & 16 & Normal & 12 & 4 & 25\% & [0, 90] \\
S16-F & 16 & Faulty & 8 & 8 & 50\% & [0, 90]; [156, 332] \\
S32-N & 32 & Normal & 16 & 16 & 50\% & [0, 90] \\
S32-F & 32 & Faulty & 24 & 8 & 25\% & [0, 90] \\
S64-N & 64 & Normal & 48 & 16 & 25\% & [0, 90]; [156, 332] \\
S64-F & 64 & Faulty & 32 & 32 & 50\% & [0, 90] \\
\bottomrule
\end{tabular}
\endgroup
\end{table*}

\subsubsection{Baselines}
The comparison includes two FDIR rules and four published software repair methods.
\begin{itemize}
\item Watchdog-FDIR checks applications in public alert order and restarts anomalous applications using the same image.
\item Table-FDIR follows maintenance dependencies and uses an event table to choose restart or a supported configuration restoration operation.
\item SWE-agent~\cite{sm_sweagent2024} uses software engineering tools to inspect sources, edit candidates, and execute tests interactively. The implementation uses its upstream interaction loop.
\item Agentless~\cite{sm_agentless2025} organizes repair into localization, candidate generation, and validation-based selection. The adapted workflow generates two candidates and executes the selected plan after ground validation.
\item RepairAgent~\cite{sm_repairagent2025} uses a finite-state machine and explicit repair hypotheses to organize evidence collection, candidate changes, and validation feedback.
\item SWE-Search~\cite{sm_swesearch2025} organizes candidate trajectories through Monte Carlo tree search, using value estimation, feedback, and candidate comparison to guide exploration.
\end{itemize}

The four published methods use the same C++ software and onboard execution interface while retaining their core control mechanisms. Common source inspection, editing, and native testing interfaces replace the language-specific and engineering tools of the original systems. State recovery and software delivery capabilities are shared. Agentless and RepairAgent use adaptations of their staged and state-guided workflows, respectively; SWE-agent uses the upstream interaction loop, and SWE-Search uses the upstream tree-search controller.

SatMend organizes related versions into joint repair sessions by component, validates each version separately, and uses the shared maintenance executor for delivery and functional verification at each target.

\subsubsection{Implementation Details}
All methods receive the same public sources, initial alerts, version inventory, service dependencies, and contact calendar. Reference implementations, private fault labels, and initialization operations are withheld from the models. All model-based methods are driven by Codex with the same model configuration and two concurrent repair slots. Builds, failed attempts, auxiliary evaluations, transfers, and activation consume the common maintenance budget. Table~\ref{tab:shared-parameters} lists model, validation, and workflow parameters; method-specific stage limits are bounded by the remaining maintenance time.

\begin{table}[!t]
\centering
\caption{Implementation parameters.}
\label{tab:shared-parameters}
\begingroup
\small
\renewcommand{\arraystretch}{1.12}
\setlength{\tabcolsep}{3pt}
\begin{tabular*}{\columnwidth}{@{\extracolsep{\fill}}lc@{}}
\toprule
\textbf{Parameter} & \textbf{Value} \\
\midrule
\multicolumn{2}{@{}l}{\textit{Shared execution and model}} \\
Concurrent repair slots & 2 \\
Regression checks & 16 per target \\
Codex model & \texttt{gpt-6.1-sol} \\
Reasoning effort / service tier & low / fast \\
\addlinespace[3pt]
\multicolumn{2}{@{}l}{\textit{Agentless}} \\
Localization limit & 45 s \\
Candidate generation limit & 65 s per candidate \\
\addlinespace[3pt]
\multicolumn{2}{@{}l}{\textit{SWE-Search}} \\
Search rounds & 12 \\
Expansions per node & 3 \\
Maximum depth & 4 \\
Candidate trajectory limit & 90 s \\
Auxiliary evaluation limit & 25 s \\
\bottomrule
\end{tabular*}
\endgroup
\end{table}

Target processes, compilation, testing, and communication execution run on a Linux host with an Intel Xeon Platinum 8358P processor and 128 logical CPUs. Event advancement handles idle waiting during disconnection, while actual model interaction and native execution times remain part of the maintenance timeline. Model-based runs are executed in batches, and timing includes the actual model service latency in each batch. All runs and failed attempts are retained.

\subsubsection{Metrics}
The primary metric is the number of initially faulty applications actually recovered before the maintenance deadline. For configuration $c$, let $\mathcal F_c$ be the initial fault set and $h_{c,a}(t)=1$ indicate that application $a$ has passed functional checks and checks of normal operation at its target. The recovery count is
\begin{equation}
N_c(t)=\sum_{a\in\mathcal F_c}h_{c,a}(t).
\label{eq:evaluation-recovery-count}
\end{equation}
The overall comparison reports counts at the deadline for each configuration and their sum over the six configurations. The recovery rate for a configuration is $N_c(t)/|\mathcal F_c|$. The contact experiment averages these rates equally across the comparison set $\mathcal C$:
\begin{equation}
\bar R(t)=\frac{1}{|\mathcal C|}
\sum_{c\in\mathcal C}\frac{N_c(t)}{|\mathcal F_c|}.
\label{eq:mean-recovery-rate}
\end{equation}
Dividing the total recovery count by 224 instead gives a proportion weighted by the number of faulty applications, which differs from this equal weighting of configurations. Passing ground tests alone does not count as recovery. A software update must complete delivery, activation, and functional verification at the actual target; state anomalies require verification after the recovery operation. Initially healthy applications also undergo checks of normal operation, and no run introduces a new functional regression.

Service metrics cover the initially affected service set $\mathcal S_c$. A service is available only when its own application and all upstream dependencies are healthy. The service recovery rate is the proportion of these services restored at 600 s. Mean service outage integrates unavailability over the maintenance interval:
\begin{equation}
D_c=\frac{1}{|\mathcal S_c|}
\int_0^{600}\sum_{s\in\mathcal S_c}
\bigl(1-A_{c,s}(t)\bigr)\,\mathrm dt,
\label{eq:mean-service-outage}
\end{equation}
where $A_{c,s}(t)$ is service availability. Metrics are first computed within each configuration and then averaged equally across configurations. Higher recovery counts and rates are better, while shorter service outages are better.

\subsection{Overall Recovery Performance}
\label{sec:overall-recovery}
This experiment evaluates batch repair performance by comparing actual application recovery within the same maintenance deadline. All seven methods use the six configurations in Table~\ref{tab:maintenance-configurations}. Table~\ref{tab:overall-recovery} groups recovery counts by satellite count and initial gateway state, with Normal and Faulty referring to the forwarding application, and reports the total over all configurations.

\begin{table}[!t]
\centering
\caption{Recovered applications within 600 s by constellation size and initial gateway state.}
\label{tab:overall-recovery}
\begingroup
\footnotesize
\renewcommand{\arraystretch}{1.22}
\setlength{\tabcolsep}{0pt}
\begin{tabular}{@{}
  >{\raggedright\arraybackslash}m{\dimexpr0.27\columnwidth-3pt\relax}
  *{2}{>{\centering\arraybackslash}m{0.10\columnwidth}}
  @{\hspace{3pt}}
  *{2}{>{\centering\arraybackslash}m{0.10\columnwidth}}
  @{\hspace{3pt}}
  *{2}{>{\centering\arraybackslash}m{0.10\columnwidth}}
  @{\hspace{3pt}}
  >{\centering\arraybackslash}m{\dimexpr0.13\columnwidth-6pt\relax}@{}}
\toprule
\multirow{2}{*}{\textbf{Method}}
& \multicolumn{2}{c}{\textbf{16 satellites}}
& \multicolumn{2}{c}{\textbf{32 satellites}}
& \multicolumn{2}{c}{\textbf{64 satellites}}
& \multirow{2}{*}{\textbf{Total}} \\
\cmidrule(l{1pt}r{2pt}){2-3}
\cmidrule(l{2pt}r{2pt}){4-5}
\cmidrule(l{2pt}r{1pt}){6-7}
& Normal & Faulty & Normal & Faulty & Normal & Faulty & \\
\midrule
Watchdog-FDIR & 2 & 0 & 8 & 0 & 8 & 0 & 18 \\
Table-FDIR & 4 & 0 & 16 & 0 & 16 & 0 & 36 \\
\addlinespace[2pt]
SWE-agent & 0 & 5 & 4 & 0 & 31 & 0 & 40 \\
Agentless & 0 & 12 & 8 & 1 & 55 & 1 & 77 \\
RepairAgent & 12 & \textbf{16} & 22 & 9 & \textbf{64} & 1 & 124 \\
SWE-Search & 5 & 9 & 21 & 0 & 63 & 17 & 115 \\
\midrule
\textbf{SatMend (ours)} & \textbf{16} & \textbf{16} & \textbf{32} & \textbf{32} & \textbf{64} & \textbf{32} & \textbf{192} \\
\bottomrule
\end{tabular}
\endgroup
\end{table}

SatMend restores 192 faulty applications, compared with 124 for RepairAgent and 115 for SWE-Search. It recovers 68 more applications than RepairAgent, the baseline with the highest total, within the same maintenance budget. Across configurations, SatMend achieves the highest or joint-highest count in all six and restores every initially faulty application in five. In S64-F, it restores 32 of 64 faulty applications; the strongest baseline in that configuration is SWE-Search, which restores 17. These results show that SatMend maintains high recovery coverage across constellation sizes and initial gateway states, translating the limited maintenance budget into recovery at more deployed targets.

\subsection{Recovery under Different Fault Compositions}
\label{sec:fault-composition}
This experiment examines recovery under mixed code faults and state anomalies by regrouping the overall results. Within each satellite count, Table~\ref{tab:fault-composition} reports recovery rates for state anomaly fractions of 25\% and 50\%. Initial gateway state, layout, and contact windows remain those of the original configurations in Table~\ref{tab:maintenance-configurations}.

\begin{table}[!t]
\centering
\caption{Recovery rates by constellation size and fraction of state anomalies.}
\label{tab:fault-composition}
\begingroup
\footnotesize
\renewcommand{\arraystretch}{1.22}
\setlength{\tabcolsep}{0pt}
\begin{tabular}{@{}
  >{\raggedright\arraybackslash}m{\dimexpr0.27\columnwidth-6pt\relax}
  *{2}{>{\centering\arraybackslash}m{0.1216667\columnwidth}}
  @{\hspace{3pt}}
  *{2}{>{\centering\arraybackslash}m{0.1216667\columnwidth}}
  @{\hspace{3pt}}
  *{2}{>{\centering\arraybackslash}m{0.1216667\columnwidth}}@{}}
\toprule
\multirow{2}{*}{\textbf{Method}}
& \multicolumn{2}{c}{\textbf{16 satellites}}
& \multicolumn{2}{c}{\textbf{32 satellites}}
& \multicolumn{2}{c}{\textbf{64 satellites}} \\
\cmidrule(l{1pt}r{2pt}){2-3}
\cmidrule(l{2pt}r{2pt}){4-5}
\cmidrule(l{2pt}r{1pt}){6-7}
& 25\% & 50\% & 25\% & 50\% & 25\% & 50\% \\
\midrule
Watchdog-FDIR & 12.50\% & 0.00\% & 0.00\% & 25.00\% & 12.50\% & 0.00\% \\
Table-FDIR & 25.00\% & 0.00\% & 0.00\% & 50.00\% & 25.00\% & 0.00\% \\
\addlinespace[2pt]
SWE-agent & 0.00\% & 31.25\% & 0.00\% & 12.50\% & 48.44\% & 0.00\% \\
Agentless & 0.00\% & 75.00\% & 3.12\% & 25.00\% & 85.94\% & 1.56\% \\
RepairAgent & 75.00\% & \textbf{100.00\%} & 28.12\% & 68.75\% & \textbf{100.00\%} & 1.56\% \\
SWE-Search & 31.25\% & 56.25\% & 0.00\% & 65.62\% & 98.44\% & 26.56\% \\
\midrule
\textbf{SatMend (ours)} & \textbf{100.00\%} & \textbf{100.00\%} & \textbf{100.00\%} & \textbf{100.00\%} & \textbf{100.00\%} & \textbf{50.00\%} \\
\bottomrule
\end{tabular}
\endgroup
\end{table}

SatMend achieves 100.00\% recovery in all three configurations with a 25\% state anomaly fraction. At 50\%, it fully restores the 16- and 32-satellite configurations and reaches 50.00\% recovery with 64 satellites, exceeding the strongest baseline in that configuration, SWE-Search, at 26.56\%. It achieves the highest or joint-highest coverage in all six size--composition combinations.

The results demonstrate the joint workflow's ability to handle both code repair and state recovery in mixed maintenance tasks. The remaining recovery gap is concentrated in the 64-satellite configuration with a 50\% state anomaly fraction; all faulty applications are restored in the other five configurations.

\subsection{Effect of Ground Contact Opportunities}
\label{sec:contact-opportunities}
\begin{figure}[!t]
  \centering
  \includegraphics[width=\linewidth]{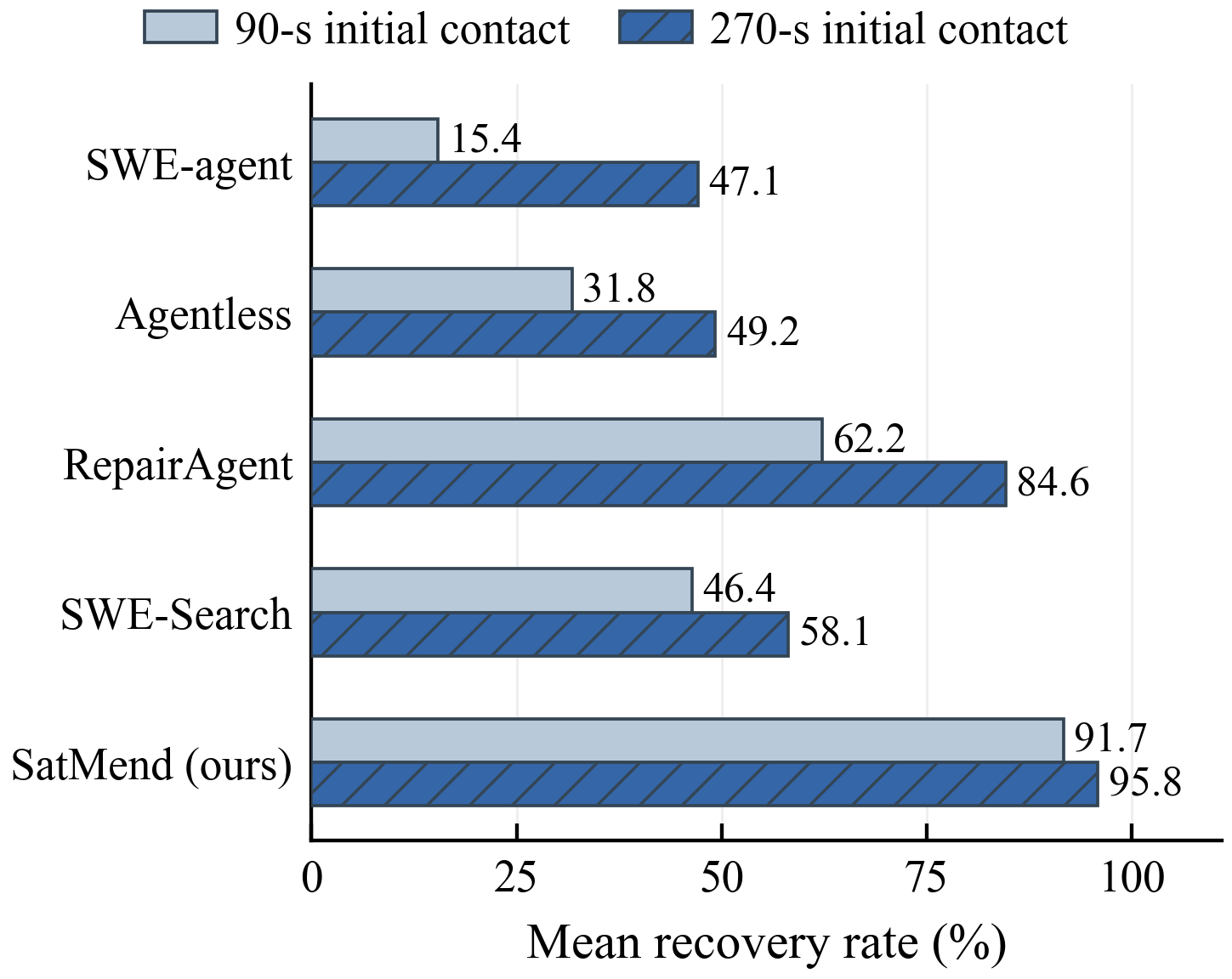}
  \caption{Mean recovery rates of fixed maintenance plans under two complete contact calendars.}
  \label{fig:contact-opportunities}
\end{figure}

This experiment evaluates how contact opportunities affect actual recovery by replaying the same repair plans under two ground contact calendars. The 30 existing plans from the five model-based methods are replayed with 90 s and 270 s of initial remaining contact, giving 60 replays. Commands, their release times, and validated binaries are fixed, as are the maintenance deadline and link capacities. Release times retain diagnosis and build delays, while diagnostic communication contention is not replayed. For configurations with a later contact window, the paired calendar moves that window from $[156,332]$ to $[336,512]$; the two calendars are therefore not nested. Fig.~\ref{fig:contact-opportunities} reports recovery rates averaged equally across configurations.

All methods recover more applications under the 270 s calendar. RepairAgent improves from 62.2\% to 84.6\%, while SatMend improves from 91.7\% to 95.8\% and retains the highest coverage under both calendars. Since neither candidates nor commands are regenerated, the additional recovery reflects the delivery opportunities required by existing plans. SatMend already completes more recovery under the shorter initial contact interval, leaving a smaller gain from additional contact.

SatMend's high coverage under the shorter initial interval indicates that its workflow makes effective use of limited contact opportunities. Timely execution of prepared repairs reduces the maintenance work left for later passes.

\subsection{Ablation of Cooperative Mechanisms}
\label{sec:ablation}
SatMend connects two designs: persistent sessions for related versions and delivery requests issued as repair progresses. The following ablations change session organization and code-delivery timing separately to examine their effects on actual recovery and maintenance cost.

\subsubsection{Joint Sessions across Versions}
This experiment evaluates how joint sessions affect completed recovery and interaction cost. On configurations with 16 satellites and a normal gateway and with 32 satellites and a faulty gateway, it compares SatMend with a variant that starts a fresh session for each version. Both use the same repair prompts, tools, model, two repair slots, and delivery executor. The per-version variant can also read public sources for other versions. Table~\ref{tab:session-ablation} reports recovery counts, workflow time, and session counts.

\begin{table}[!t]
\centering
\caption{Ablation of joint repair sessions across versions.}
\label{tab:session-ablation}
\begingroup
\small
\renewcommand{\arraystretch}{1.20}
\setlength{\tabcolsep}{3pt}
\begin{tabular}{@{}
  >{\raggedright\arraybackslash}m{\dimexpr0.34\columnwidth-1.5\tabcolsep\relax}
  >{\centering\arraybackslash}m{\dimexpr0.23\columnwidth-1.5\tabcolsep\relax}
  >{\centering\arraybackslash}m{\dimexpr0.25\columnwidth-1.5\tabcolsep\relax}
  >{\centering\arraybackslash}m{\dimexpr0.18\columnwidth-1.5\tabcolsep\relax}@{}}
\toprule
\textbf{Method}
& \shortstack{\textbf{Recovered}\\\textbf{applications}}
& \shortstack{\textbf{Workflow}\\\textbf{time (s)}}
& \shortstack{\textbf{Model}\\\textbf{sessions}} \\
\midrule
\multicolumn{4}{@{}l}{\textit{16 satellites; normal gateway}} \\
\addlinespace[2pt]
Per-version sessions & 16 & 144.9 & 6 \\
\textbf{SatMend (ours)} & 16 & \textbf{81.6} & \textbf{2} \\
\midrule
\multicolumn{4}{@{}l}{\textit{32 satellites; faulty gateway}} \\
\addlinespace[2pt]
Per-version sessions & 15 & 177.2 & 8 \\
\textbf{SatMend (ours)} & \textbf{32} & \textbf{111.4} & \textbf{2} \\
\bottomrule
\end{tabular}
\endgroup
\end{table}

In the 16-satellite configuration, both variants restore all 16 faulty applications. Joint sessions reduce workflow time from 144.9 s to 81.6 s and the session count from six to two, reducing the interaction cost of completing the same task. In the 32-satellite configuration with a faulty gateway, joint sessions reduce time from 177.2 s to 111.4 s, increase recovery from 15 to 32 applications, and reduce sessions from eight to two.

The two configurations show complementary effects of session organization: lower workflow cost when full recovery is achievable, and greater completed recovery when execution opportunities are limited. The ablation supports processing related versions continuously within component sessions, preserving shared diagnostic context and reducing the need to reorganize maintenance for each version.

\subsubsection{Code-Delivery Timing}
\begin{figure}[!t]
  \centering
  \includegraphics[width=\linewidth]{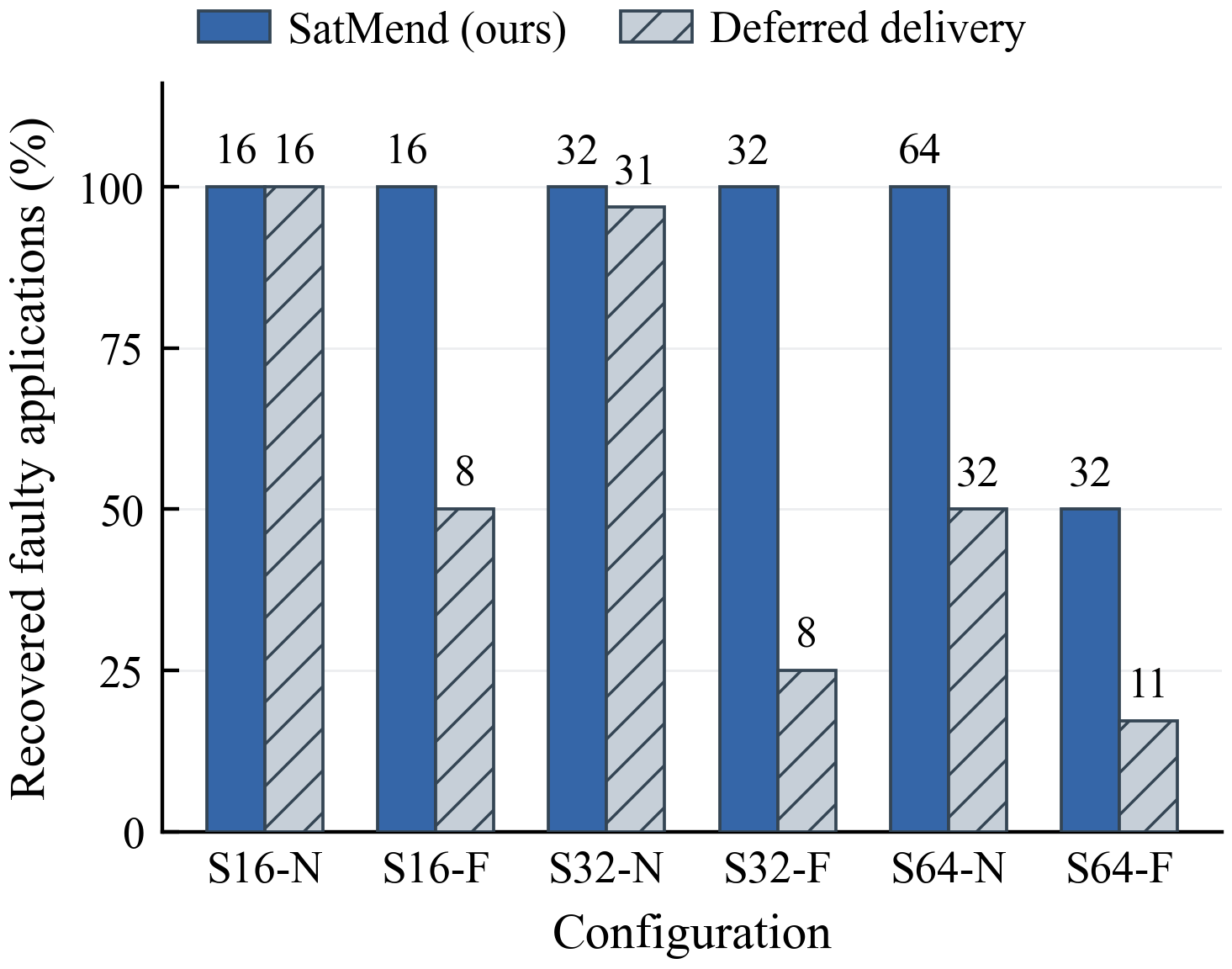}
  \caption{Ablation of code-delivery timing with fixed repair artifacts.}
  \label{fig:delivery-ablation}
\end{figure}

This experiment evaluates delivery timing by delaying code deployment while fixing SatMend's binaries, state operations, and retries in all six primary configurations. Code deployments are released only after the last candidate evaluation in the corresponding run; state-operation times and contact calendars remain unchanged. Fig.~\ref{fig:delivery-ablation} compares actual recovery under the original and delayed delivery times.

Delaying delivery reduces recovery in five of the six configurations. S32-F falls from 32 recovered applications to eight, and S64-N from 64 to 32. S16-N still achieves full recovery, indicating that the effect depends on repair progress and available contact. Where recovery declines, releasing identical candidates later prevents the original recovery outcome, demonstrating the execution constraints between ground validation and onboard recovery.

The differences with fixed repair artifacts show that timely release of validated code candidates increases actual recovery before the deadline. Together, the two ablations support the complementary roles of SatMend's designs: joint sessions reduce repeated interaction and maintenance cost, while delivery during continued repair allows prepared candidates to use contact windows and restore their targets.

\subsection{Maintenance Case Study}
\label{sec:maintenance-case}
\begin{figure}[!t]
  \centering
  \includegraphics[width=\linewidth]{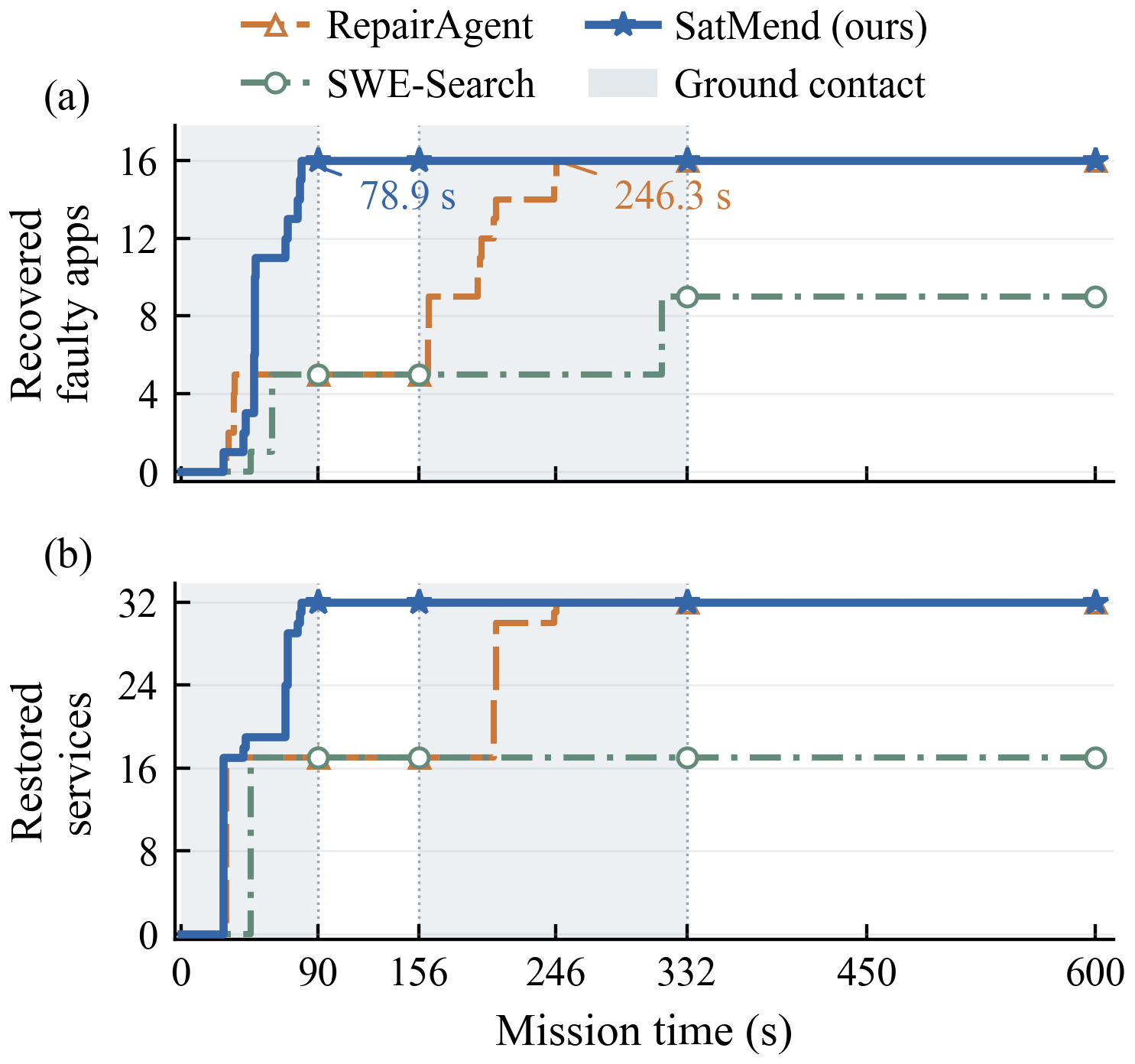}
  \caption{Application and service recovery in S16-F.}
  \label{fig:maintenance-case}
\end{figure}

This case examines how gateway recovery, candidate preparation, and delivery combine to restore services across the satellites. S16-F contains 32 applications on 16 satellites, with 16 direct anomalies and 32 initially affected services. Its contact windows are $[0,90]$ and $[156,332]$. Fig.~\ref{fig:maintenance-case} compares application and service recovery for SatMend, RepairAgent, and SWE-Search on a common timeline. Table~\ref{tab:maintenance-milestones} lists gateway recovery, validation of all code candidates, and full recovery.

\begin{table}[!t]
\centering
\caption{Maintenance milestones in S16-F.}
\label{tab:maintenance-milestones}
\begingroup
\small
\renewcommand{\arraystretch}{1.22}
\setlength{\tabcolsep}{3pt}
\begin{tabular}{@{}
  >{\raggedright\arraybackslash}m{\dimexpr0.32\columnwidth-1.5\tabcolsep\relax}
  >{\centering\arraybackslash}m{\dimexpr0.20\columnwidth-1.5\tabcolsep\relax}
  >{\centering\arraybackslash}m{\dimexpr0.23\columnwidth-1.5\tabcolsep\relax}
  >{\centering\arraybackslash}m{\dimexpr0.25\columnwidth-1.5\tabcolsep\relax}@{}}
\toprule
\textbf{Method}
& \shortstack{\textbf{Gateway}\\\textbf{recovered}\\\textbf{(s)}}
& \shortstack{\textbf{All code}\\\textbf{candidates}\\\textbf{validated (s)}}
& \shortstack{\textbf{Full}\\\textbf{recovery}\\\textbf{(s)}} \\
\midrule
RepairAgent & 29.8 & 232.9 & 246.3 \\
SWE-Search & 45.5 & 551.8 & Not achieved \\
\midrule
\textbf{SatMend (ours)} & \textbf{27.9} & \textbf{54.8} & \textbf{78.9} \\
\bottomrule
\end{tabular}
\endgroup
\end{table}

All three methods restore the gateway early: SatMend at 27.9 s, RepairAgent at 29.8 s, and SWE-Search at 45.5 s. Their later differences therefore concern whether candidate preparation and target operations continue effectively once the gateway is available. SatMend validates candidates for all four code implementations by 54.8 s and completes target verification for all faulty applications by 78.9 s. Both recovery trajectories reach their upper limits before the first contact interval ends at 90 s.

RepairAgent validates all four code candidates only at 232.9 s and achieves full recovery at 246.3 s during the second contact interval. SWE-Search reaches the candidate-validation milestone at 551.8 s, after the final contact window within the deadline, and leaves applications and services unrecovered. The two recovery trajectories show how component progress affects the restoration of dependent services: early gateway repair alone does not ensure timely completion of the maintenance task.

This case connects gateway recovery, candidate preparation, and target delivery. SatMend restores the maintenance path early and completes preparation and delivery for the remaining versions within the first contact interval, allowing component recovery to restore dependent services promptly.

\section{Conclusion}
This paper presented SatMend, a ground-based LLM-agent method for software repair in satellite constellations with intermittent ground contact. We formulated maintenance as maximizing verified target recovery before a common deadline under software-version, dependency, and contact constraints. SatMend retains diagnostic context across related versions in persistent sessions, uses source and runtime evidence to guide code repair and state recovery, and coordinates validation for each version with timely delivery to deployed targets. A software-in-the-loop evaluation across 16--64-satellite configurations shows that SatMend restores 192 of 224 faulty applications within 600~s, compared with 124 for RepairAgent, the strongest baseline in aggregate. Ablations support the complementary roles of joint sessions and timely delivery, while the case study connects early repair preparation to application and dependent-service recovery within the first contact interval. These results demonstrate the effectiveness of coordinating ground repair preparation with onboard execution to restore more applications and their services under intermittent contact. Future work will explore adaptive session allocation and dependency-aware delivery prioritization for changing fault batches and contact opportunities.

\bibliographystyle{IEEEtran}
\bibliography{ref}

\onecolumn
\appendices

\section{Prompt Templates and Recorded Interaction}
\label{app:prompts}
Each repair slot receives a task description, its assigned versions and applications, the public inventory, and the remaining maintenance budget. The task description is followed by instructions for handling the assigned versions jointly. The two boxes below reproduce these instructions, with field headings added for readability and instance-specific entries represented by placeholders. Tool descriptions and argument schemas are supplied through the tool interface. Subsequent tool results return to the same Codex session; the controller does not insert a separate fixed prompt after each call.

\subsection{Initial Task and Joint Repair Instructions}
The prompts use the following abbreviations: low Earth orbit (LEO), central processing unit (CPU), ground--satellite link (GSL), and inter-satellite link (ISL).

\begin{prompttemplate}{Initial maintenance task}
\textbf{Mission and resources.} You are ground maintenance for a cooperative LEO satellite software cluster. Only the exposed tools may be used. No shell, private files, acceptance references, or other models. All compilation, native applications and native tests run on the separate CPU host. The ground Codex is the only repair model. There is one shared real mission deadline, two model work slots, finite GSL contact, and a shared ISL. S0's independent updater remains available; its relay app can block downstream maintenance. The common executor delivers real executable bytes and verifies their actual native behavior. Code generation, failed attempts and any helper model calls cost time.

\textbf{Diagnosis and recovery.} Restore actual native applications and the dependent services before the mission deadline. A service alarm can be caused by its prerequisite, a component code error, or runtime state. Inspect native evidence and public source before deciding what requires a code change. A restart or supported configuration restoration can recover some conditions without editing code. These public actions are identical for all methods. Native tests report aggregate operation success; the public component contract and interface headers provide the required semantics.

\textbf{Source and validation constraints.} Every deployed variant, including healthy comparison instances, has a ground production archive. Only source files explicitly declared editable can be changed. Test and driver code is immutable. Unknown code errors have no precomputed rollback patch. Reuse reasoning where justified by source and version, but validate each candidate variant independently. A passing ground test alone is not deployed recovery. Every compatible instance must actually receive/activate or recover.

\textbf{Disconnected operation.} Do useful ground analysis when links are unavailable; do not repeatedly poll an unavailable path.

\smallskip\noindent Assigned variants: \emph{[source-version identifiers assigned to this slot]}

\smallskip\noindent Assigned apps: \emph{[application identifiers associated with those versions]}

\smallskip\noindent Public inventory: \emph{[deployment inventory; cached service alerts; editable files; component contracts; supported recovery actions; service dependencies; contact calendar]}

\smallskip\noindent Remaining real budget: \emph{[remaining time in seconds]}
\end{prompttemplate}

\begin{prompttemplate}{Instructions appended for a SatMend session}
Handle the assigned variants jointly. Start from public dependency/reachability relations and native observations to distinguish local defects from propagated alarms. Compare related source versions and runtime evidence before editing. Reuse causal analysis for the shared component; prepare all required changes for its complete contract, then validate EACH variant independently. Use documented state recovery instead of code changes when appropriate. Coordinate source repair with actual delivery to all assigned compatible applications. Prioritize a maintenance prerequisite when evidence justifies it. Other worker(s) own the remaining variants, which you may read and inspect for context but must not edit. Finish after feasible assigned repairs have been delivered.
\end{prompttemplate}

\subsection{Tool Interface and Returned Evidence}
The interface separates ground preparation from operations that require contact with deployed targets. Source inspection, editing, and native candidate validation can proceed without a ground link. Target diagnostics, recovery commands, and receipt retrieval depend on the available maintenance path. Table~\ref{tab:maintenance-tools} summarizes the exposed operations. Ground validation returns build diagnostics and aggregate functional results; target-facing calls additionally report delivery status or received execution evidence.

\begin{table}[!htbp]
\centering
\caption{Tools available to a SatMend session.}
\label{tab:maintenance-tools}
\begingroup
\small
\renewcommand{\arraystretch}{1.22}
\setlength{\tabcolsep}{5pt}
\begin{tabular}{@{}
  >{\raggedright\arraybackslash}p{\dimexpr0.25\linewidth-1.333333\tabcolsep\relax}
  >{\raggedright\arraybackslash}p{\dimexpr0.32\linewidth-1.333333\tabcolsep\relax}
  >{\raggedright\arraybackslash}p{\dimexpr0.43\linewidth-1.333333\tabcolsep\relax}@{}}
\toprule
\textbf{Operation} & \textbf{Purpose} & \textbf{Returned evidence} \\
\midrule
\texttt{overview} & Read the public maintenance context & Inventory, cached observations, current time, and contact schedule \\
\texttt{list\_files}, \texttt{read\_file}, \texttt{search} & Inspect ground sources and public interfaces & Editable-file lists, source text, and matching locations \\
\texttt{replace\_text} & Edit an assigned production source & Edit result and digest of the updated source contents \\
\texttt{evaluate} & Build and test one source version & Build result, aggregate functional checks, and source digest \\
\texttt{inspect} & Diagnose a deployed application & Received image identifier, functional results, telemetry, and dependency status, or a path failure \\
\texttt{recover} & Restart the same image or restore a supported configuration & Received execution and target-verification results, or a delivery failure \\
\texttt{deploy} & Validate and submit a candidate to compatible targets & Validation result and per-target delivery or queue status \\
\texttt{deployment\_status} & Retrieve target execution receipts & Received transfer, activation, and functional-verification records \\
\texttt{finish} & Submit the session summary & Session submission acknowledgement \\
\bottomrule
\end{tabular}
\endgroup
\end{table}

\FloatBarrier
\subsection{Recorded Interaction in S16-F}
The following example condenses one recorded FileManager session in S16-F. The symbols $v_A$ and $v_B$ replace the two source-version identifiers; application identifiers are retained. Each entry reports selected calls and returned evidence in their recorded order. Source reads and edits between the entries are summarized.

\begin{prompttemplate}{Abridged interaction: gateway repair and downstream delivery}
\textbf{1. Inspect the maintenance path.}
\texttt{inspect(app=``S0\_relay'')} reports one successful operation out of five, with service unavailable. A later \texttt{inspect(app=``S13\_1'')} returns \texttt{forwarding\_unavailable}.

\smallskip\noindent\textbf{2. Prepare and validate both versions.}
After reading the two FileManager sources and their public file-system interface, the agent edits both assigned versions. Separate \texttt{evaluate} calls for $v_A$ and $v_B$ each pass all five functional operations and the checks of normal operation.

\smallskip\noindent\textbf{3. Submit the gateway candidate.}
\texttt{deploy(variant=}$v_A$\texttt{, apps=[``S0\_relay''])} returns a queued operation. The session continues reading the other assigned components, then requests \texttt{deployment\_status(app=``S0\_relay'')}. The returned receipt records completed activation and successful target verification at 27.9~s.

\smallskip\noindent\textbf{4. Reassess downstream alarms.}
After gateway recovery, \texttt{inspect(app=``S2\_0'')} reports nominal functionality and available service. \texttt{inspect(app=``S13\_1'')} remains at one successful operation out of five. These returned observations distinguish a propagated interruption from a remaining local defect.

\smallskip\noindent\textbf{5. Deliver the second version.}
\texttt{deploy(variant=}$v_B$\texttt{, apps=[``S13\_1'', ``S4\_0''])} queues both targets. Later \texttt{deployment\_status} calls return successful target verification at 40.8 and 42.1~s, respectively. Subsequent inspections report all five operations passing and service available at both targets.
\end{prompttemplate}

The two versions use the same component context but retain separate validation records and delivery requests. Restoring the gateway first makes downstream diagnosis and execution possible; subsequent observations identify which applications still need their own repair. The session ends with a summary of the completed work, while recovery accounting uses the target verification records.

\FloatBarrier
\section{Dependent Service Recovery and Remaining Faults}
\label{app:service-recovery}
This experiment examines whether application recovery also restores services blocked by failed dependencies and identifies the maintenance stages at which remaining faults persist. The six configurations from Section~\ref{sec:experimental-setup} are grouped by initial gateway state. All methods use the same service dependency graphs, and Table~\ref{tab:service-recovery} compares affected-service recovery and mean outage over 0--600 s.

\begin{table}[!htbp]
\centering
\caption{Service recovery under different initial gateway states.}
\label{tab:service-recovery}
\begingroup
\footnotesize
\renewcommand{\arraystretch}{1.22}
\setlength{\tabcolsep}{0pt}
\begin{tabular}{@{}
  >{\raggedright\arraybackslash}m{\dimexpr0.28\columnwidth-3pt\relax}
  *{2}{>{\centering\arraybackslash}m{0.18\columnwidth}}
  @{\hspace{3pt}}
  *{2}{>{\centering\arraybackslash}m{0.18\columnwidth}}@{}}
\toprule
\multirow{2}{*}{\textbf{Method}}
& \multicolumn{2}{c}{\textbf{Initially normal gateway}}
& \multicolumn{2}{c}{\textbf{Initially faulty gateway}} \\
\cmidrule(l{1pt}r{2pt}){2-3}
\cmidrule(l{2pt}r{1pt}){4-5}
& \shortstack{\textbf{Recovery}\\\textbf{rate}}
& \shortstack{\textbf{Mean}\\\textbf{outage (s)}}
& \shortstack{\textbf{Recovery}\\\textbf{rate}}
& \shortstack{\textbf{Mean}\\\textbf{outage (s)}} \\
\midrule
Watchdog-FDIR & 0.00\% & 600.0 & 0.00\% & 600.0 \\
Table-FDIR & 0.00\% & 600.0 & 0.00\% & 600.0 \\
\addlinespace[2pt]
SWE-agent & 16.15\% & 536.2 & 17.71\% & 522.8 \\
Agentless & 28.65\% & 487.6 & 51.82\% & 329.3 \\
RepairAgent & 64.58\% & 271.5 & 71.61\% & 221.1 \\
SWE-Search & 40.10\% & 444.3 & 34.64\% & 406.4 \\
\midrule
\textbf{SatMend (ours)} & \textbf{100.00\%} & \textbf{84.8} & \textbf{91.67\%} & \textbf{94.2} \\
\bottomrule
\end{tabular}
\endgroup
\end{table}

SatMend achieves the highest service recovery rate and shortest mean outage in both groups. With a normal gateway, it restores all affected services with a mean outage of 84.8 s. With a faulty gateway, it restores 91.67\% with a mean outage of 94.2 s, compared with 71.61\% and 221.1 s for RepairAgent. SatMend thus restores more affected services and reduces their time without service. Although the two FDIR rules recover some applications in the main comparison, neither restores affected services, indicating that failed dependencies still block those services.

\begin{figure}[!htbp]
  \centering
  \includegraphics[width=\linewidth]{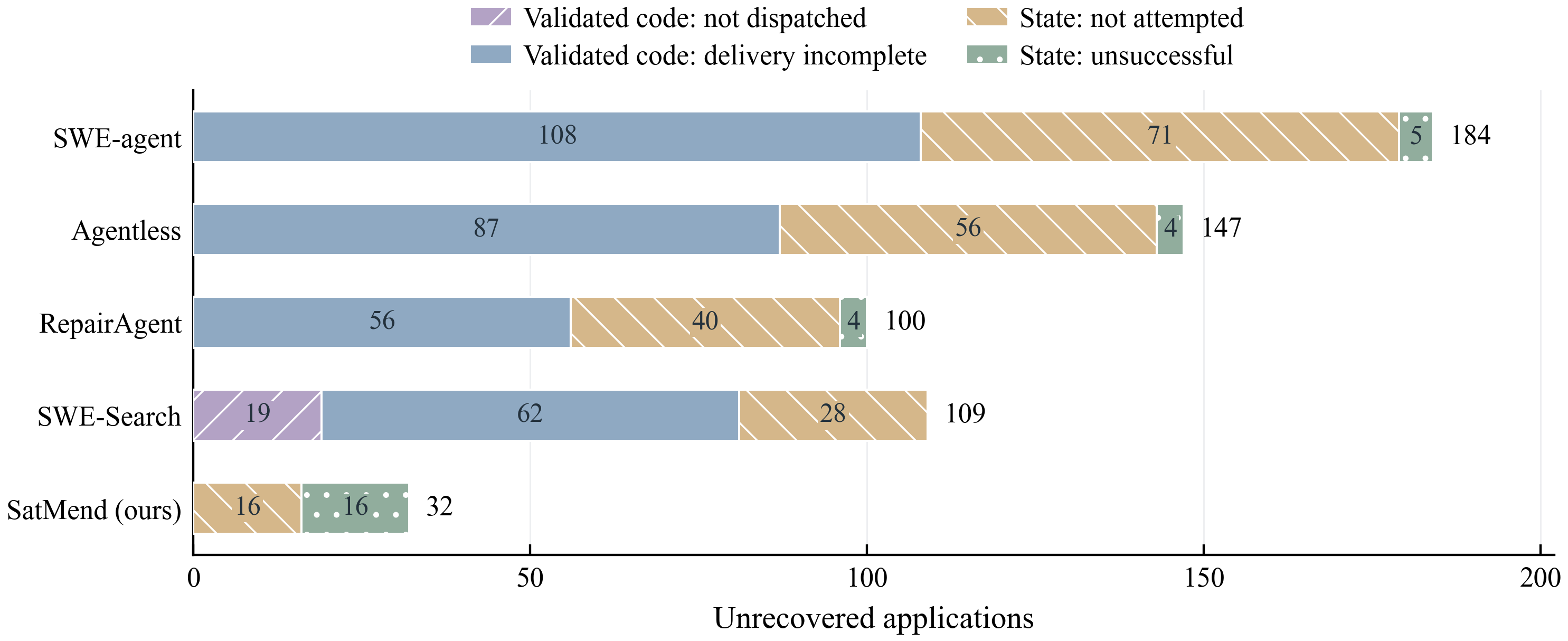}
  \caption{Maintenance stages of unrecovered applications.}
  \label{fig:remaining-faults}
\end{figure}

To examine the remaining interruption, unrecovered applications at 600 s are classified by maintenance stage for the five model-based methods. Code cases distinguish whether an available candidate has been dispatched and whether delivery has completed; state cases distinguish whether a recovery operation has been attempted. Fig.~\ref{fig:remaining-faults} accounts for all 572 unrecovered instances across these methods.

All remaining code cases have validated candidates that are either not dispatched or have incomplete delivery. RepairAgent has 56 such instances; SWE-Search has 62 with incomplete delivery and 19 not dispatched. SatMend has no remaining code cases. Its 32 unrecovered applications all have state anomalies: no recovery operation is attempted for 16, and attempted operations fail to restore the other 16. Of the initial 84 state-anomaly applications, SWE-Search restores 56, compared with 52 for SatMend. These results locate SatMend's main gain in applying code repairs at their targets, while also identifying a remaining gap in state recovery.

Table~\ref{tab:service-recovery} and Fig.~\ref{fig:remaining-faults} connect service outcomes to component maintenance progress. Ground-validated candidates must complete delivery and target verification before affected dependent services can resume. SatMend delivers and verifies repairs for all code-fault targets and achieves broader, earlier service recovery, extending the benefit of batch repair to dependent services.

\FloatBarrier
\section{Recovery Progress and Runtime Cost}
\label{app:recovery-cost}
Recovery counts at the deadline measure completed work, while recovery trajectories show when those gains become available. Workflow and model costs quantify the effort required. This experiment combines recovery timelines and interaction records from the six configurations to assess timeliness and efficiency.

\subsection{Time to Recovery}
\begin{figure}[!htbp]
  \centering
  \includegraphics[width=\linewidth]{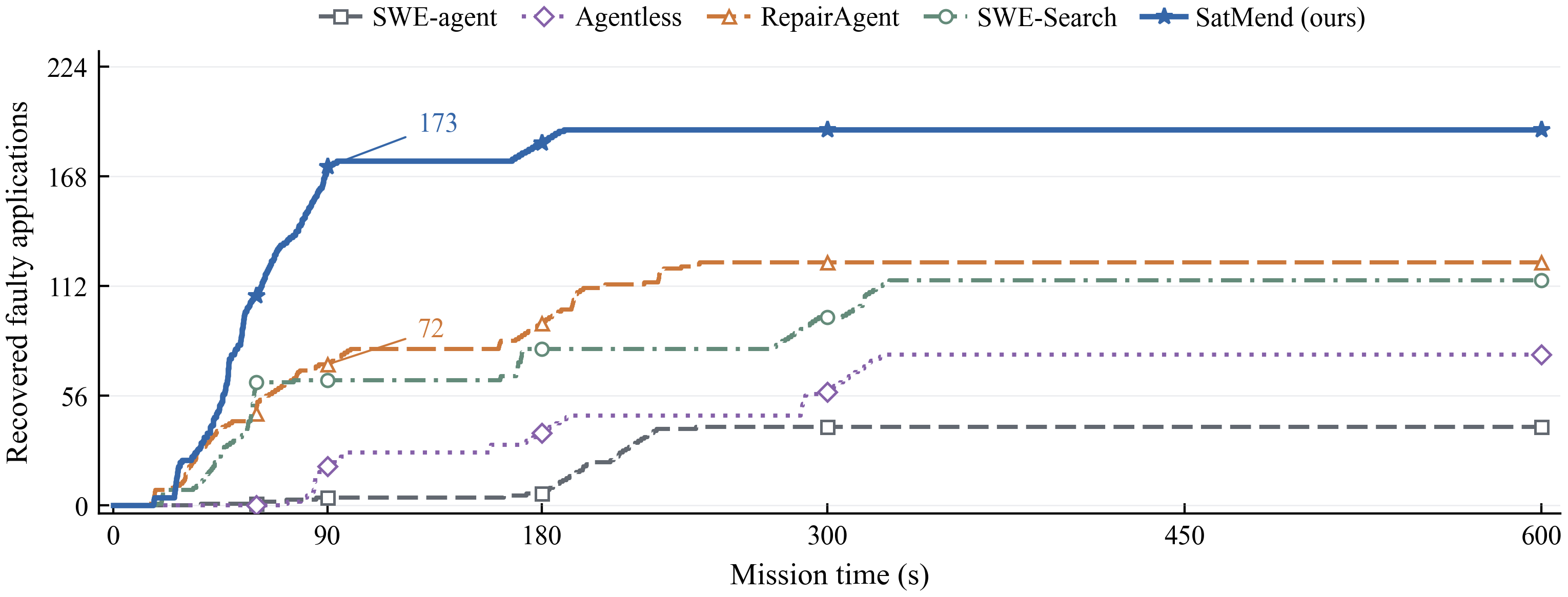}
  \caption{Cumulative actual recovery across the six configurations.}
  \label{fig:recovery-progress}
\end{figure}

This experiment compares cumulative recovery trajectories for the five model-based methods. Counts sum the initially faulty applications that have passed verification at their actual targets across all six configurations. Fig.~\ref{fig:recovery-progress} follows recovery through the 600 s maintenance interval; its endpoints match the recovery totals in Table~\ref{tab:overall-recovery}.

At 90 s, when the initial contact interval ends, SatMend has restored 173 faulty applications, compared with 72 for RepairAgent. Most of SatMend's eventual recovery is already available by this point, with the count later increasing to 192. Baseline recoveries extend further into the maintenance interval and still end at lower totals. The trajectories thus reveal a timing advantage beyond the final counts: more SatMend targets return to operation early in maintenance.

Together with the contact replays in Section~\ref{sec:contact-opportunities}, this early recovery shows that SatMend applies more repairs during the initial contact interval, leaving fewer tasks for later passes. Its advantage concerns both the number of restored applications and the time at which they resume operation.

\FloatBarrier
\subsection{Workflow and Model Cost}
This experiment measures workflow time and model usage for the five model-based methods. Workflow time is averaged across the six configurations, while model sessions, uncached input tokens, and output tokens are summed. Auxiliary evaluations, failed attempts, and retries are included. In Table~\ref{tab:workflow-cost}, $\geq$ indicates a reported lower bound because some sessions did not return complete usage records.

\begin{table}[!htbp]
\centering
\caption{Workflow and model costs across six configurations.}
\label{tab:workflow-cost}
\begingroup
\footnotesize
\renewcommand{\arraystretch}{1.22}
\setlength{\tabcolsep}{2pt}
\begin{tabular}{@{}
  >{\raggedright\arraybackslash}m{\dimexpr0.28\columnwidth-1.6\tabcolsep\relax}
  >{\centering\arraybackslash}m{\dimexpr0.16\columnwidth-1.6\tabcolsep\relax}
  >{\centering\arraybackslash}m{\dimexpr0.14\columnwidth-1.6\tabcolsep\relax}
  >{\centering\arraybackslash}m{\dimexpr0.23\columnwidth-1.6\tabcolsep\relax}
  >{\centering\arraybackslash}m{\dimexpr0.19\columnwidth-1.6\tabcolsep\relax}@{}}
\toprule
\textbf{Method}
& \shortstack{\textbf{Mean}\\\textbf{workflow}\\\textbf{time (s)}}
& \shortstack{\textbf{Model}\\\textbf{sessions}}
& \shortstack{\textbf{Uncached}\\\textbf{input tokens}\\($10^6$)}
& \shortstack{\textbf{Output}\\\textbf{tokens}\\($10^3$)} \\
\midrule
SWE-agent & 452.7 & 430 & $\geq 10.872$ & $\geq 37.7$ \\
Agentless & 395.0 & 168 & $\geq 5.403$ & $\geq 114.2$ \\
RepairAgent & 197.7 & 42 & 1.889 & 40.5 \\
SWE-Search & 509.1 & 260 & $\geq 6.658$ & $\geq 144.8$ \\
\midrule
\textbf{SatMend (ours)} & \textbf{119.5} & \textbf{12} & \textbf{0.681} & \textbf{22.5} \\
\bottomrule
\end{tabular}
\endgroup
\end{table}

SatMend uses 12 model sessions across the six configurations, with a mean workflow time of 119.5 s; RepairAgent uses 42 sessions and averages 197.7 s. Their uncached inputs are 0.681 and 1.889 million tokens, respectively, and their outputs are 22.5 and 40.5 thousand tokens. Together with the overall recovery results in Table~\ref{tab:overall-recovery}, these results show that SatMend restores more applications with fewer sessions and lower input and output usage.

The costs also reflect the interaction demands of different control workflows. SWE-agent repeatedly issues single-action requests with interaction history, while SWE-Search additionally performs search expansions and auxiliary evaluations. SatMend retains persistent component sessions to organize maintenance across multiple versions. The joint-session ablation in Section~\ref{sec:ablation} shows a corresponding reduction in session count and workflow time. Together, the progress and cost results indicate that session organization and timely delivery convert more of the limited maintenance budget into actual target recovery.

\end{document}